\documentclass[aps,prl,superscriptaddress,twocolumn,longbibliography]{revtex4-1}
\usepackage{bbm}
\usepackage{mathrsfs}
\usepackage{amsmath}
\usepackage{amsfonts}
\usepackage[colorlinks=true,linkcolor=blue,urlcolor=blue,citecolor=blue,anchorcolor=blue]{hyperref}
\usepackage{graphicx,epstopdf}
\usepackage{subfigure}
\usepackage{epsfig}
\usepackage{dcolumn}
\usepackage{bm}
\usepackage{color}
\usepackage{natbib}
\usepackage{amssymb}
\usepackage{xcolor}
\usepackage{braket}
\usepackage{ulem}
\usepackage{float}
\usepackage{braket}
\usepackage{changes}
\allowdisplaybreaks[4]

\begin{document}
	
\title{Universal classification of continuous phase transitions with two order parameters}
\author{Yilun Xu}
\email{xuyilun@alumni.pku.edu.cn}
\affiliation{State Key Laboratory for Mesoscopic Physics, School of Physics, Frontiers Science Center for Nano-optoelectronics, Peking University, Beijing 100871, China}
\affiliation{Beijing Academy of Quantum Information Sciences, Beijing 100193, China}
\author{Feng-Xiao~Sun}

\email{sunfengxiao@bupt.edu.cn}
\affiliation{State Key Laboratory of Information Photonics and Optical Communications $\&$ School of Physical Science and Technology, Beijing University of Posts and Telecommunications, Beijing 100876, China}

\begin{abstract}
	The continuous phase transition, indicated by the macroscopic order parameter and the occurrence of the spontaneous symmetry breaking, is well illustrated based on the Ginzburg-Landau's paradigm. In systems described by one order parameter, the phase diagram is only composed of the normal phase and the ordered phase. However, in systems with two or more order parameters, much richer phase diagrams and critical phenomena may emerge. According to coupling terms of the free energy, we classify the systems with two order parameters into several categories, featuring different patterns of phase diagrams and critical scalings respectively. Our work propose the new avenue to studying and evaluating the complex systems only based on the coupling terms in expressions of free energy.
	
	
	\end{abstract} 
	
	\maketitle
	
	\textit{Introduction.} 
	As the one of the most fundamental theories in condensed matter physics, Ginzburg–Landau's theory (also called Landau's theory)~\cite{HOHENBERG20151} clarifies that the birth of macroscopic order parameters indicates the spontaneous symmetry breaking, which is well-known as the phase transition. Although the Landau's theory, which is based on mean-field approximation, ignores the higher order correlations and the quantum fluctuations, it has long been a powerful tool to depict the systems in thermodynamic limit or classical oscillator limit where the classic property dominates the status of system. It explains not only the continuous symmetry breaking, such as the $U(1)$ symmetry breaking in the superconductivity~\cite{PhysRev.108.1175}, superfluidity of $^4\rm He$~\cite{RN97,RN105,RN94,RN104,RN103}, and the $O(3)$ symmetry breaking in Heisenberg model~\cite{PhysRevLett.20.589,RN98}, but also the discrete symmetric systems, including $Z_2$ symmetry breaking in classic Ising ferromagnetic model~\cite{Peierls1936}, quantum Rabi model~\cite{PhysRevLett.115.180404,PhysRevLett.119.220601} and Dicke model~\cite{PhysRevA.7.831,PhysRevE.67.066203,HEPP1973360,PhysRevLett.104.130401,PhysRevLett.107.140402,PhysRevLett.112.086401,PhysRevLett.120.183603,PhysRevLett.122.193605}.

	According to the Landau's theory, the free energy is written as the function of the order parameters. The system status is determined by the minimum position of the free energy. While the change of environmental factors reshapes the expression of free energy, leading to the new system status. In general, the system described by one order parameter will give us only the normal and ordered phase, seperated by a series of critical environmental parameters $\vec{\lambda}_c\equiv\left(\lambda^1_c,\lambda^2_c,\dots\right)^T$. The minimum points of free energy in the two phases are featured by the zero and nonzero order parameter respectively. However, the situations will become much more complex with two or more ordes involved. For example, in optical lattice systems or cavity-mediated quantum materials, the photon-induced long range potential will compete with the short-range interaction of the atomic ensemble, which potenitally manifests the inteplays of different orders, such as the superradiance and the superfluid~\cite{PhysRevA.100.013611,PhysRevA.104.053313,xu2026superradianceenhancessuppressesfermionic}, the magnetic orders and superradiance~\cite{PhysRevLett.128.080601,PhysRevA.98.043613,rao2025unilateralcriticalityphasetransition}, or the density wave and the pairing order~\cite{PhysRevA.96.051602}. The coexistences and competitions of multi-order parameters improve our knowledge on conventional physical models and benefit us to searching for exotic matter phases and developing novel quantum materials~\cite{PhysRevLett.132.073602,PhysRevLett.84.4068,PhysRevA.104.053313,PhysRevLett.125.053602,PhysRevLett.127.177002,PhysRevLett.130.083603,PhysRevA.87.023831,PhysRevB.103.075131}.
	
	Given the complicacy and diversity, it's always an attractive topic to extract the universal regularity in such systems. We find that all the complex phase diagrams and scaling behaviors reveal general features which can be summarized from the coupling terms of the free energy. Therefore, we are trying to construct the universal framework for all such systems with all phase transitions and interplays of different orders well categorized, which will provide useful instructions for further investigates on similar systems.

	In this letter, concentrating on continuous phase transition in systems with two order parameters, we first classify the typical patterns  of phase diagram into different categories based on general free energy expression of two order parameters. These typical patterns are viewed as the basic structure of the universal phase diagram by considering the fragments, defections and the splices. Besides, we give the general critical behaviors on all kinds of phase boundaries in each category. In order to study the feasibility of our results, two concrete physical systems are taken as examples. For the quantum Rabi model in classical oscillator limit and the Dicke dimer in thermodynamic limit, the phase diagrams and the scaling behaviors are well illustrated by our proposal. It's convincing that our result leads to a general avenue to dealing with complicated systems and analyze novel physical phenomena.
	
	\begin{figure*}[tb]
		\centering
		\includegraphics[width=1.0\textwidth]{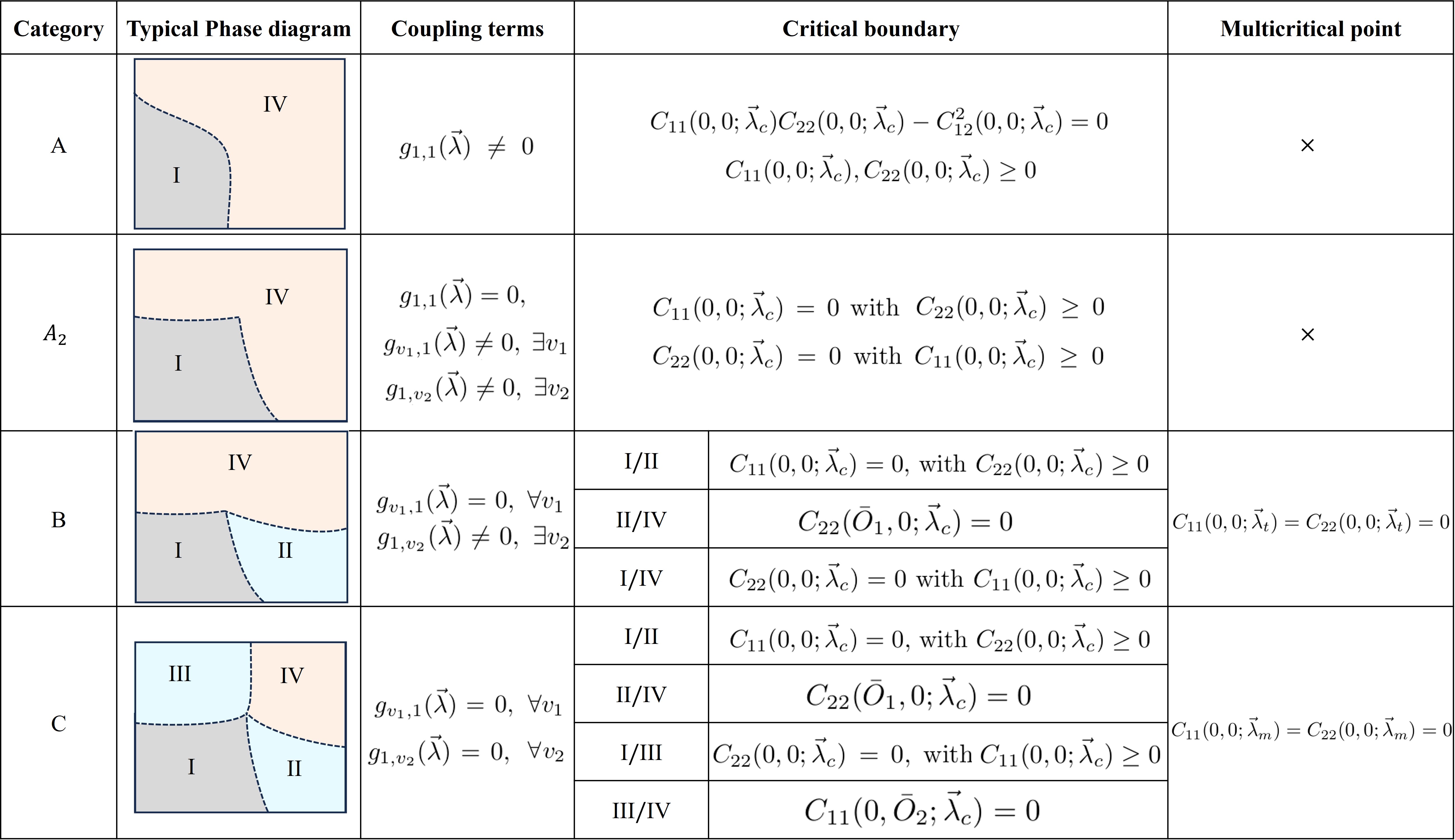}
		\caption{The universal categories of the phase diagrams are summarized in the table. The typical patterns, expressions of the free energy, critical boundaries and the multicritical points are displayed in the 2nd, 3rd, 4th and 5th column respectively.}\label{main_t1}
	\end{figure*}
	
	\textit{Universal categories of systems with two order parameters.} Consider the free energy of a system is described by two macroscopic order parameters $\left(O_1,O_2\right)$ with the environmental parameter vector $\vec{\lambda}$, expressed as
	\begin{align}\label{free_energy_main}
		F=F(O_1,O_2;\vec{\lambda})=\sum_{v_1,v_2}g_{v_1,v_2}(\vec{\lambda})O_1^{v_1}O_2^{v_2}.
	\end{align}
	Generally, the order parameter $O_i$ is well selected as nonzero real numbers, such as the number of photon in the superradiant systems or the magnetism in ferromagnetic materials. 
	
The system status will be determined by the order parameters $\left(\bar{O}_1(\vec{\lambda}),\bar{O}_2(\vec{\lambda})\right)$, where the free energy is minimum, i.e., it satisfies
	\begin{align}
		F(\bar{O}_1(\vec{\lambda}),\bar{O}_2(\vec{\lambda}))=\min_{(O_1,O_2)}F(O_1,O_2;\vec{\lambda}).
	\end{align}
	 Notice that the system status $\left(\bar{O}_1(\vec{\lambda}),\bar{O}_2(\vec{\lambda})\right)$ is the functions of the environmental parameter $\vec{\lambda}$. The phases with order parameter $\bar{O}_i=0$ and $\bar{O}_i\neq0$ represent the normal phase and the ordered phase respectively, which are seperated by the critical point $\vec{\lambda}_c$. 
	 
	 Given that the general free energy~(\ref{free_energy_main}) is rather complex for us to determine the system status and critical boundaries, we give a few assumptions before further discussions. On one hand, we consider the linear term is absent, where $g_{1,0}(\vec{\lambda})=g_{0,1}(\vec{\lambda})=0,~\forall\vec{\lambda}$, in order to ganrantee the existence of normal phase $\bar{O}_i=0$. On the other hand, we assume the lowest order terms is quadratic. In other words, the situation $g_{2,0}(\vec{\lambda})g_{0,2}(\vec{\lambda})=0,~\forall\vec{\lambda}$ is excluded. Besides, we assume that all the coefficients are the continuous and analytical function against the environmental parameter $\vec{\lambda}$. Additionally, we only concentrate on the continuous phase transition, whose critical boundaries and scaling rates are analytically accessible. 
	 
	 Follow the former paragraph, all possible statuses are sorted into four different kinds, $\left(\bar{O}_1=0,\bar{O}_2=0\right)$, $\left(\bar{O}_1\neq0,\bar{O}_2=0\right)$, $\left(\bar{O}_1=0,\bar{O}_2\neq0\right)$ and $\left(\bar{O}_1\neq0,\bar{O}_2\neq0\right)$, marked by region I, II, III and IV respectively. However, not all these four phases exist as the "stable phase" in every system. All the systems are further categorized according to the expressions of free energy~(\ref{free_energy_main}), and only the specific phases and patterns of phase diagram are available in each category. 
	 
	 All the situations are divided into four categories as A, $A_2$, B and C shown in Fig.\ref{main_t1}. The typical phase diagrams are displaced in the second column, right behind which, the coupling terms, boundaries are multi-critical points are listed as well. In the phase diagram, the grey, blue and orange region stand for the phases I, II(III) and IV respectively. And the black solid lines represent the continuous phase boundaries. The category A and $A_2$ only support the phase I and IV. While they are distinguished by whether the coupling term $g_{1,1}O_1O_2$ exists in the free energy, leading to different critical scalings which will be discussed later. The Category B supports phase I, II and IV. The existence of phase II asks for a region satisfying $\dfrac{\partial F}{\partial O_2}\Bigg|_{O_2=0}=0$, implying all the terms $g_{v_1,1}O_1^{v_1}O_2$ are absent. The category with I, III and IV supported are also viewed as category B if we exchange the subscripts 1 and 2. As the last case, all the four regions might emergy in Category C, with both terms like $g_{v_1,1}O_1^{v_1}O_2$ and terms $g_{1,v_2}O_1O_2^{v_2}$ absent.
	 
	 According to the Landau's theory, the disappearance of the 2nd order differentiations indicates the critical boundaries of the continuous phase transitions, hence we give all the critical conditions in the third column, where we define $C_{ij}\equiv\partial^2F/\partial O_i\partial O_j$ to simplify the writing. In the Category B and C with more than two phases coexisting, the multicritical points might emerge in the intersections of the critical boundaries. Detailed proofs for the critical boundaries and multicritical points are available in~\cite{supplementary}, where we also demonstrate the phase diagrams by numerically solving several self-defined expressions of free energy.
	 

	 \begin{figure}[tb]
	 	\centering
	 	\hspace{-0.6cm}
	 	\includegraphics[width=0.5\textwidth]{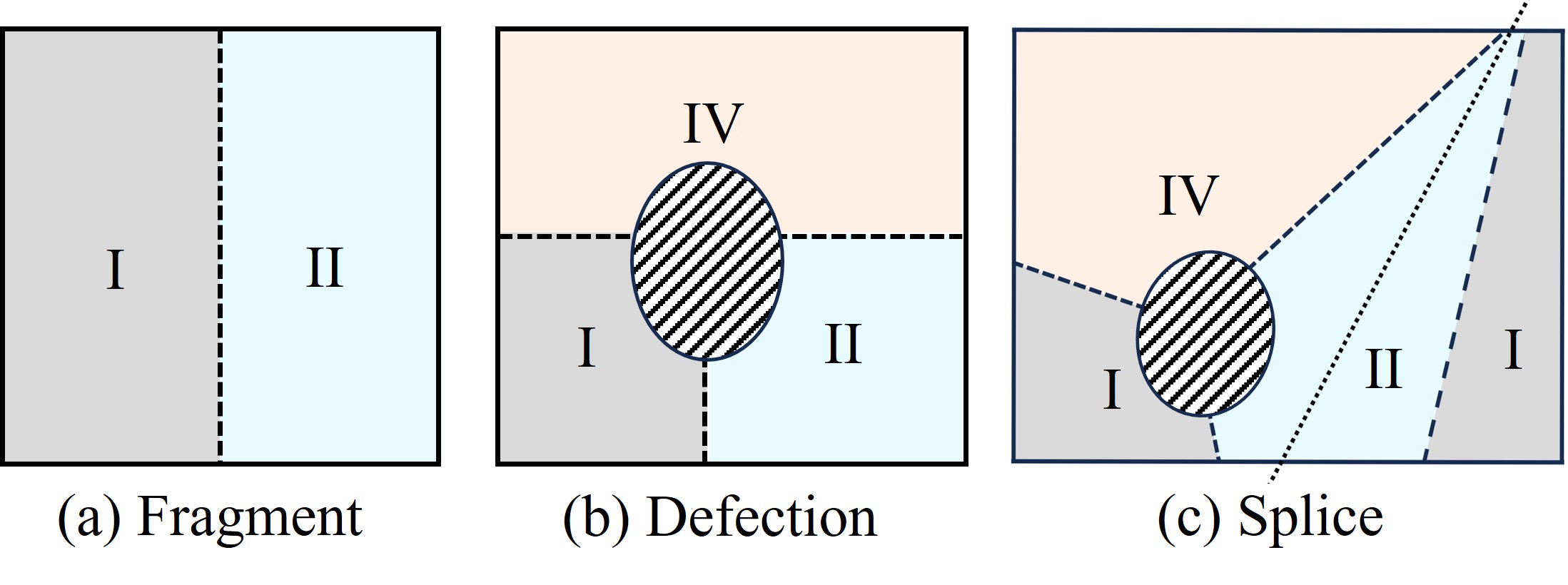}
	 	\caption{The three general configurations of the real phase diagrams. The shadows represent the defections in the continuous phase diagrams.}\label{main_real}
	 \end{figure}
	 
	 In Fig.~\ref{main_t1}, we only demonstrate the typical patterns of phase diagram in each case. However, the real phase diagrams might be more complex, and the value ranges of the function cluster $\left\{g_{v_1,v_2}(\vec{\lambda})\right\}$ decide the real structure of phase diagram. Take Category B as the example, we divide all general situations into three kinds shown in (a-c) of Fig.~\ref{main_real}. 
	 	
	 The first kind is named as fragment, where not all phases are available. As is shown in subfigure (a), the phase IV is absent here. The second kind is the defection, referring to the region broken by the 1st order phase transition. The subfigure (b) demonstrates the example that the defection swallows the junction among the three phases. In addition, the splices are also possible as the combinations of several parts, such as the diagram in (c), which can be viewed as the combination of the fragment diagram (a) and the defection diagram (b). Apart from the simple examples listed here, detailed discussions and phase diagrams are contained in~\cite{supplementary}.
	 
	 \begin{figure*}[tb]
	 	\centering
	 	\includegraphics[width=1.0\textwidth]{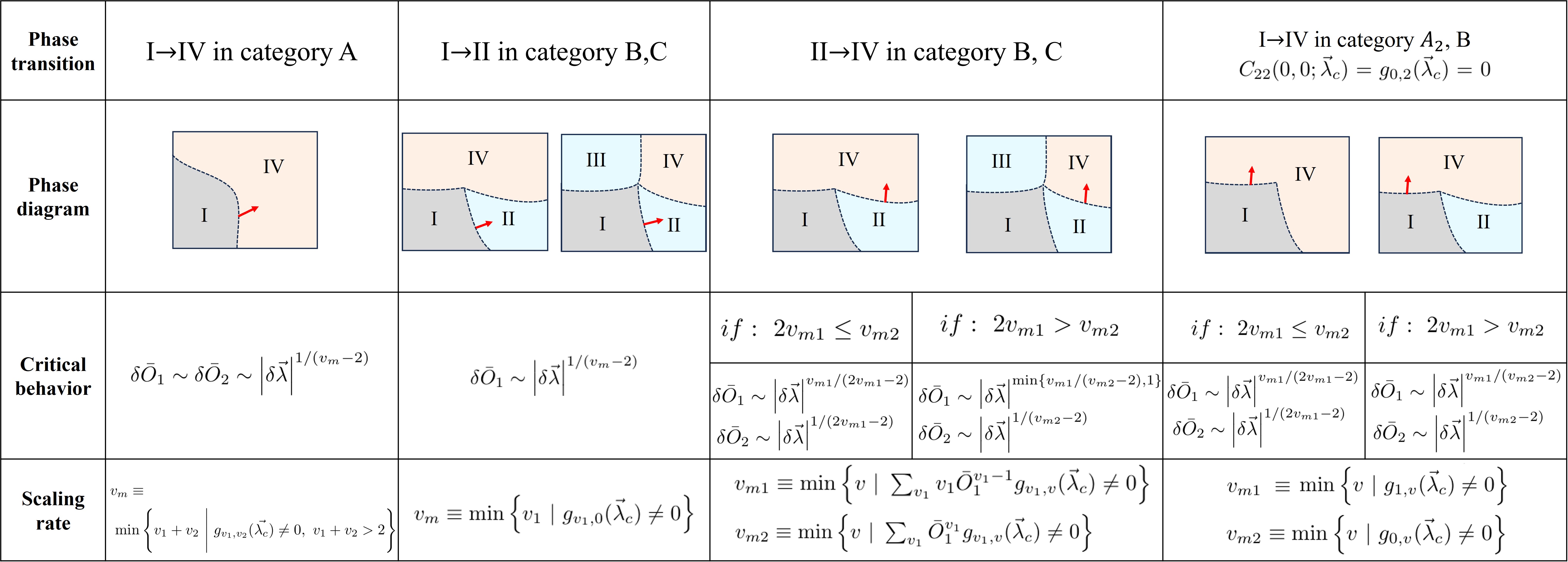}
	 	\caption{The several different critical behaviors are listed in this table, where the 3rd row gives the critical behaviors for the two orders and the 4th row shows the corresponding scaling rates.}\label{main_t2}
	 \end{figure*}
	 
	 \textit{Critical behaviors.} Around the phase boundary, the critical scalings determine the evoluting tendency of order parameters, which further influence the property in the whole phase. We display all possible scaling behaviors in Fig.~\ref{main_t2}, divided into four different kinds. The first two kinds give the same scaling law with different definition $v_m$. For the transition I$\to$IV in Category A, the order parameter $O_1$ and $O_2$ emerges simultaneously, $v_m$ is defined as the lowest order of the coupling term $O_1^{v_1}O_2^{v_2}$ besides the quadratic term $O_1O_2$. While the transition I$\to$II in Category B and C, $v_m$ is defined as the lowest order of the term $O_1^{v_1}$. The other two kinds give more complex scaling behaviors, where the transition of $O_2$ will influence the value of $O_1$ indirectly, leading to the interplay of different physcial quantities. As a special case of transition II$\to$IV, where $\partial_1F(\bar{O}_1,0;\vec{\lambda}_c+\delta\vec{\lambda})=0$, the changing rates between $\delta \bar{O}_1$ and $\delta \bar{O}_2$ is given as
	 \begin{align}\label{interplay}
	 	\delta \bar{O}_1\approx-\dfrac{\partial_1\partial_2^{v_{m1}}F(\bar{O}_1,0;\vec{\lambda}_c)}{v_{m1}!C_{11}(\bar{O}_1,0;\vec{\lambda}_c)}(\delta \bar{O}_2)^{v_{m1}},
	 \end{align}
	 which is the same as that in transition I$\to$IV of category $A_2$ and B~\cite{supplementary}. This scaling relations provide the reference to evaluate the property of the phase only based on the differentiation at the critical point, the detailed practice of which is illustrated in the accompanying manuscript~\cite{xu2026superradianceenhancessuppressesfermionic}.
	 
	 In the following, we review several examples, combining the universal phase diagrams and critical scalings proposed here.

\textit{Phase transition in quantum Rabi model.} As the first example, we introduce the quantum Rabi model. The Hamiltonian reads $H_{\rm Rabi}=\omega a^\dagger a+\dfrac{\Omega}{2}\sigma_z+\lambda(a+a^\dagger)\sigma_x$.
Distinguished from the previous researches~\cite{PhysRevLett.115.180404}, we define two order parameters $O_1\equiv\alpha$ and $O_2\equiv\left<\sigma_x\right>$ here. In classic oscillator limit, where the atomic transition frequency $\Omega$ is much larger than the cavity frequency $\omega$, both the higher order correlations and the quantum fluctuation are suppressed by the ratio $\Omega/\omega\to\infty$, thus the mean-field theory can be safely applied. The atomic part is almost separable with the classical optical mode, giving $\left<(a+a^\dagger)\sigma_x\right>\approx\left<a+a^\dagger\right>\left<\sigma_x\right>$. Also, the optical mode $a$ is replaced by the displacement $\alpha$ as $a\rightarrow\alpha\equiv\left<a\right>\sim\sqrt{\Omega/\omega}$. And the superradiant phase transition occurs after the critical point $\lambda_c=\sqrt{\Omega\omega}/2$~\cite{PhysRevLett.115.180404}. In closed system, the $\alpha$ is a real number, and the ground state energy reads
\begin{align}
	E(\alpha,\left<\sigma_x\right>)=\omega\alpha^2-\dfrac{\Omega}{2}(1-\left<\sigma_x\right>^2)^{1/2}+2\lambda\alpha\left<\sigma_x\right>,
\end{align}
where we apply the spin conservation rule $\left<\sigma_z\right>^2+\left<\sigma_x\right>^2+\left<\sigma_y\right>^2=1$ with $\left<\sigma_y\right>=0$. After expanding the energy function, we find the expression matches well with the Category A type, indicating the $\left<\sigma_x\right>$ and $\left<a\right>$ will be transferred into ordered simultaneouly. The Hessian matrix is given as
 \begin{align}
 	\mathcal{H}=\left(
	\begin{array}{cc}
		\omega&\lambda\\
		\lambda&\Omega/4
	\end{array}\right),
\end{align}
which give the critical boundary $\lambda_c^2=\Omega\omega/4$, matches well with the previous results. The displacement in the superradiant phase gives $\left|\alpha\right|=\sqrt{(\Omega/16g^2\omega)(16g^4-1)}\sim (g-g_c)^{1/2}$ as $g\to g_c$, where $g\equiv\lambda/\sqrt{\Omega\omega}$ is the dimensionless coupling strength, and $g_c=1/2$ is the critical point of $g$. The $1/2$ scaling rate refers to $v_m=4$ in our criterion, which can be checked by expanding $(1-\left<\sigma_x\right>^2)^{1/2}$ till the second order term. Once the system enters into the superradiant region, the order parameter $\left<\sigma_x\right>=\bra{\tilde{\downarrow}}\sigma_x\ket{\tilde{\downarrow}}$, where $\ket{\tilde{\downarrow}}\equiv-\sin\theta\ket{\uparrow}+\cos\theta\ket{\downarrow}$ with $\tan(2\theta)=4\lambda\alpha/\Omega$. Thus we can get $\left<\sigma_x\right>\sim\left|\alpha\right|$ near the critical point $g\to g_c$. 

\begin{figure}[tb]
	\centering
	\hspace{-0.6cm}
	\includegraphics[width=0.51\textwidth]{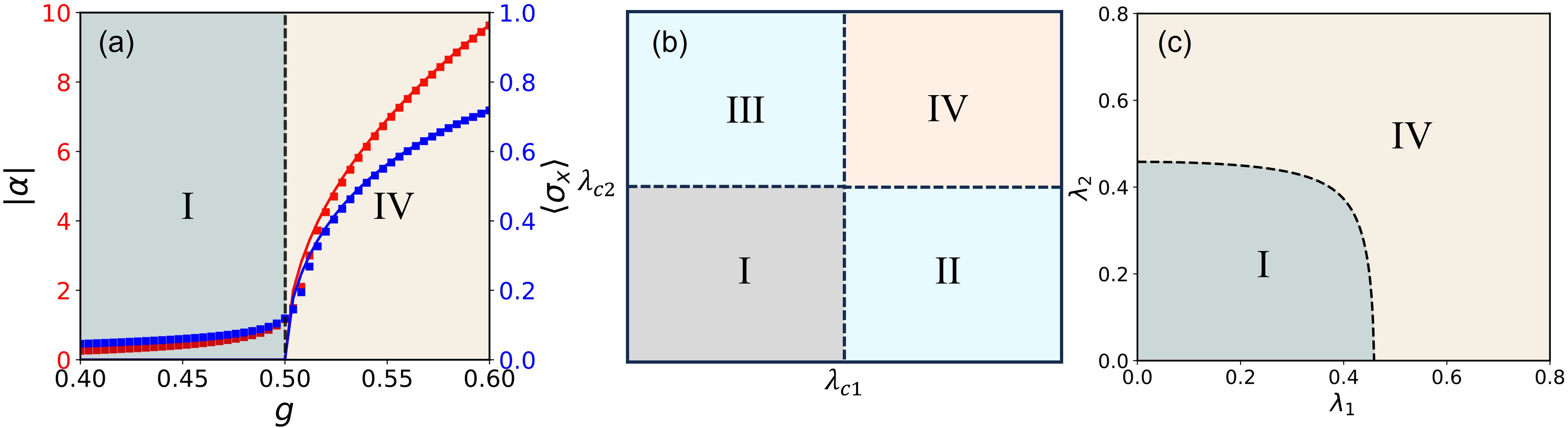}
	\caption{(a) The results of the order parameters $\left|\alpha\right|$ and $\left<\sigma_x\right>$ in quantum Rabi model are demonstrated by red and blue lines respectively. We also give the phase diagram in $J=0$ and $J=0.2$ cases of asymmetric Dicke dimer in (b) and (c) respectively. (b)(c) are also seen in the supplementary material of~\cite{PhysRevLett.133.233604} with the same parameters.}\label{dimer}
\end{figure}

The analytical and numerical results are given in Fig.~\ref{dimer}(a) with fixed ratio $\Omega/\omega=500$. The solid and dotted lines stands for the analytical and numerical results respectively, which match well with each other, verifying our classification of Categroy A and the corresponding scaling behavior.

\textit{Asymmetric Dicke dimer.} Consider the case where two asymmetric Dicke model coupled through the photonic hopping term with $H_{\rm hop}=J(a_1+a_1^\dagger)(a_2+a_2^\dagger)$. The Hamiltonian of the system can be expressed as $H_{\rm sys}=\sum_{j=1,2}H_j^{\rm Dicke}+H_{\rm hop}$, where the Dicke Hamiltonian of cavity $j=1,2$ reads $H_j^{\rm Dicke}=\omega_{cj}a_j^\dagger a_j+\omega_{aj} S_j^z+{2\lambda_j}S_j^x(a_j+a_j^\dagger)/{\sqrt{N_j}}$~\cite{PhysRevLett.133.233604}.


In thermodynamic limit $N_j\to\infty$, each single Dicke model is in the normal (superradiant) phase if $\lambda_j < \lambda_{cj}$ ($\lambda_j > \lambda_{cj}$) respectively. where the critical coupling strength is given by mean-field approximation as~\cite{PhysRevLett.112.173601,PhysRevLett.120.183603}
	 \begin{equation}
	 	\lambda_{cj}= \frac{1}{2}\sqrt{\omega_{aj}(\omega_{cj}+{\kappa_j^2}/{\omega_{cj}})}\,
	 \end{equation}
	 with $\kappa_j$ the respective cavity decay rate. And $\kappa_j\to0$ gives us the critical boundary for ground state in closed system.
	 
If we choose the order parameter $O_j=\left<a_j\right>/\sqrt{N_j}$ respectively, and choose $\lambda_j$ as the environmental parameters, we can predict the phase diagrams of the dimer system according to the categorizations. Consider the special case that the hopping rate J vanishes, the Hamiltonian shows the formula of category C. Obviously, the system will be reduced into two isolated Dicke model in this case, and the phase diagram is divided into four regions by $\lambda_{cj}$ as in Fig.~\ref{dimer}(b), matching well with general features of the category C. On the other hand, for a nonzero hopping rate $J$, the Hamiltonian gives the form as category A, with the critical point $(J_c,\lambda_{c1},\lambda_{c2})$ is satisfies
\begin{align}
	J_c=2\sqrt{\left(\lambda_{c1}^2-\lambda_1^2\right) \left(\lambda_{c2}^2-\lambda_2^2\right) /(\omega_{a1} \omega_{a2})  },
\end{align}
in parametric region. And the analytical phase diagram only contains two regions $NP\&NP$ and $SRP\&SRP$ shown in (c), validating our categorization.

Consider the special symmetric case where $\lambda_1=\lambda_2$, the order parameters are analytically accessible, giving $O_1=-O_2={\rm Re}\gamma^e_{a}=\pm\frac{1}{4\lambda}\sqrt{(\frac{1}{4(Z^e_{a})^2}-1)}$, and $Z^e_{a}=\frac{1}{8\lambda^2}(2J-1-\kappa^2)$, with fixed repulsive hopping rate $J>0$. Therefore, we obtain the same scaling rate $v_m=4$ as $\delta O_j\sim[\lambda-\lambda_c(J)]^{1/2}$ when $\lambda\to\lambda_c(J)$, where $\lambda_c(J)$ is the critical coupling strength with hopping rate $J$, and we set $\omega_c=\omega_a=1$ as the unit.

\textit{Summary.} Both the classical or quantum phase transitions have been widely studied during the past several decades, the generalization to two or more order parameters system is still lack of knowledge though. In this work, we give the univeral phase diagrams and critical scalings of continuous phase transition in two order parameters case in Landau's paradigm. Although the quantum fluctuation is ignored based on mean-field approximation, it still serves as a good approximation in classic oscillator limit or thermodynamic limit. Based on these results, the previous models can be mapped into our classifications. Consider the quantum Rabi model in classical oscillator limit and the Dicke dimer in thermodynamic limit as two typical example, it turns out the phase diagrams, critical boundaries and the scaling rates are well explained by means of our approach.


Our work proposes a universal scheme to study a general system with two physcial quantities. The expression of the coupling terms provides the precise physical insights on the typical shape of phase diagrams and the critical scalings. Our proposal can also extended to systems with more order parameters, and it facilitates us to understand complex physical phenomena from a higher prespective.


\begin{acknowledgments}
	This work is supported by the National Natural Science Foundation of China (Grants No. 12474256) and the Quantum Science and Technology-National Science and Technology Major Project (Grant No. 2025ZD0301000).
\end{acknowledgments}

		\bibliography{ref}

\begin{thebibliography}{37}%
\makeatletter
\providecommand \@ifxundefined [1]{%
 \@ifx{#1\undefined}
}%
\providecommand \@ifnum [1]{%
 \ifnum #1\expandafter \@firstoftwo
 \else \expandafter \@secondoftwo
 \fi
}%
\providecommand \@ifx [1]{%
 \ifx #1\expandafter \@firstoftwo
 \else \expandafter \@secondoftwo
 \fi
}%
\providecommand \natexlab [1]{#1}%
\providecommand \enquote  [1]{``#1''}%
\providecommand \bibnamefont  [1]{#1}%
\providecommand \bibfnamefont [1]{#1}%
\providecommand \citenamefont [1]{#1}%
\providecommand \href@noop [0]{\@secondoftwo}%
\providecommand \href [0]{\begingroup \@sanitize@url \@href}%
\providecommand \@href[1]{\@@startlink{#1}\@@href}%
\providecommand \@@href[1]{\endgroup#1\@@endlink}%
\providecommand \@sanitize@url [0]{\catcode `\\12\catcode `\$12\catcode
  `\&12\catcode `\#12\catcode `\^12\catcode `\_12\catcode `\%12\relax}%
\providecommand \@@startlink[1]{}%
\providecommand \@@endlink[0]{}%
\providecommand \url  [0]{\begingroup\@sanitize@url \@url }%
\providecommand \@url [1]{\endgroup\@href {#1}{\urlprefix }}%
\providecommand \urlprefix  [0]{URL }%
\providecommand \Eprint [0]{\href }%
\providecommand \doibase [0]{http://dx.doi.org/}%
\providecommand \selectlanguage [0]{\@gobble}%
\providecommand \bibinfo  [0]{\@secondoftwo}%
\providecommand \bibfield  [0]{\@secondoftwo}%
\providecommand \translation [1]{[#1]}%
\providecommand \BibitemOpen [0]{}%
\providecommand \bibitemStop [0]{}%
\providecommand \bibitemNoStop [0]{.\EOS\space}%
\providecommand \EOS [0]{\spacefactor3000\relax}%
\providecommand \BibitemShut  [1]{\csname bibitem#1\endcsname}%
\let\auto@bib@innerbib\@empty
\bibitem [{\citenamefont {Hohenberg}\ and\ \citenamefont
  {Krekhov}(2015)}]{HOHENBERG20151}%
  \BibitemOpen
  \bibfield  {author} {\bibinfo {author} {\bibfnamefont {P.}~\bibnamefont
  {Hohenberg}}\ and\ \bibinfo {author} {\bibfnamefont {A.}~\bibnamefont
  {Krekhov}},\ }\href {\doibase https://doi.org/10.1016/j.physrep.2015.01.001}
  {\bibfield  {journal} {\bibinfo  {journal} {Physics Reports}\ }\textbf
  {\bibinfo {volume} {572}},\ \bibinfo {pages} {1} (\bibinfo {year} {2015})},\
  \bibinfo {note} {an introduction to the Ginzburg–Landau theory of phase
  transitions and nonequilibrium patterns}\BibitemShut {NoStop}%
\bibitem [{\citenamefont {Bardeen}\ \emph {et~al.}(1957)\citenamefont
  {Bardeen}, \citenamefont {Cooper},\ and\ \citenamefont
  {Schrieffer}}]{PhysRev.108.1175}%
  \BibitemOpen
  \bibfield  {author} {\bibinfo {author} {\bibfnamefont {J.}~\bibnamefont
  {Bardeen}}, \bibinfo {author} {\bibfnamefont {L.~N.}\ \bibnamefont {Cooper}},
  \ and\ \bibinfo {author} {\bibfnamefont {J.~R.}\ \bibnamefont {Schrieffer}},\
  }\href {\doibase 10.1103/PhysRev.108.1175} {\bibfield  {journal} {\bibinfo
  {journal} {Phys. Rev.}\ }\textbf {\bibinfo {volume} {108}},\ \bibinfo {pages}
  {1175} (\bibinfo {year} {1957})}\BibitemShut {NoStop}%
\bibitem [{\citenamefont {Kapitza}(1938)}]{RN97}%
  \BibitemOpen
  \bibfield  {author} {\bibinfo {author} {\bibfnamefont {P.}~\bibnamefont
  {Kapitza}},\ }\href {\doibase 10.1038/141074a0} {\bibfield  {journal}
  {\bibinfo  {journal} {Nature}\ }\textbf {\bibinfo {volume} {141}},\ \bibinfo
  {pages} {74} (\bibinfo {year} {1938})}\BibitemShut {NoStop}%
\bibitem [{\citenamefont {Greiner}\ \emph {et~al.}(2002)\citenamefont
  {Greiner}, \citenamefont {Mandel}, \citenamefont {Esslinger}, \citenamefont
  {Hänsch},\ and\ \citenamefont {Bloch}}]{RN105}%
  \BibitemOpen
  \bibfield  {author} {\bibinfo {author} {\bibfnamefont {M.}~\bibnamefont
  {Greiner}}, \bibinfo {author} {\bibfnamefont {O.}~\bibnamefont {Mandel}},
  \bibinfo {author} {\bibfnamefont {T.}~\bibnamefont {Esslinger}}, \bibinfo
  {author} {\bibfnamefont {T.~W.}\ \bibnamefont {Hänsch}}, \ and\ \bibinfo
  {author} {\bibfnamefont {I.}~\bibnamefont {Bloch}},\ }\href {\doibase
  10.1038/415039a} {\bibfield  {journal} {\bibinfo  {journal} {Nature}\
  }\textbf {\bibinfo {volume} {415}},\ \bibinfo {pages} {39} (\bibinfo {year}
  {2002})}\BibitemShut {NoStop}%
\bibitem [{\citenamefont {Bigagli}\ \emph {et~al.}(2024)\citenamefont
  {Bigagli}, \citenamefont {Yuan}, \citenamefont {Zhang}, \citenamefont
  {Bulatovic}, \citenamefont {Karman}, \citenamefont {Stevenson},\ and\
  \citenamefont {Will}}]{RN94}%
  \BibitemOpen
  \bibfield  {author} {\bibinfo {author} {\bibfnamefont {N.}~\bibnamefont
  {Bigagli}}, \bibinfo {author} {\bibfnamefont {W.}~\bibnamefont {Yuan}},
  \bibinfo {author} {\bibfnamefont {S.}~\bibnamefont {Zhang}}, \bibinfo
  {author} {\bibfnamefont {B.}~\bibnamefont {Bulatovic}}, \bibinfo {author}
  {\bibfnamefont {T.}~\bibnamefont {Karman}}, \bibinfo {author} {\bibfnamefont
  {I.}~\bibnamefont {Stevenson}}, \ and\ \bibinfo {author} {\bibfnamefont
  {S.}~\bibnamefont {Will}},\ }\href {\doibase 10.1038/s41586-024-07492-z}
  {\bibfield  {journal} {\bibinfo  {journal} {Nature}\ }\textbf {\bibinfo
  {volume} {631}},\ \bibinfo {pages} {289} (\bibinfo {year}
  {2024})}\BibitemShut {NoStop}%
\bibitem [{\citenamefont {Gopalakrishnan}\ \emph {et~al.}(2009)\citenamefont
  {Gopalakrishnan}, \citenamefont {Lev},\ and\ \citenamefont
  {Goldbart}}]{RN104}%
  \BibitemOpen
  \bibfield  {author} {\bibinfo {author} {\bibfnamefont {S.}~\bibnamefont
  {Gopalakrishnan}}, \bibinfo {author} {\bibfnamefont {B.~L.}\ \bibnamefont
  {Lev}}, \ and\ \bibinfo {author} {\bibfnamefont {P.~M.}\ \bibnamefont
  {Goldbart}},\ }\href {\doibase 10.1038/nphys1403} {\bibfield  {journal}
  {\bibinfo  {journal} {Nat. Phys.}\ }\textbf {\bibinfo {volume} {5}},\
  \bibinfo {pages} {845} (\bibinfo {year} {2009})}\BibitemShut {NoStop}%
\bibitem [{\citenamefont {Baumann}\ \emph {et~al.}(2010)\citenamefont
  {Baumann}, \citenamefont {Guerlin}, \citenamefont {Brennecke},\ and\
  \citenamefont {Esslinger}}]{RN103}%
  \BibitemOpen
  \bibfield  {author} {\bibinfo {author} {\bibfnamefont {K.}~\bibnamefont
  {Baumann}}, \bibinfo {author} {\bibfnamefont {C.}~\bibnamefont {Guerlin}},
  \bibinfo {author} {\bibfnamefont {F.}~\bibnamefont {Brennecke}}, \ and\
  \bibinfo {author} {\bibfnamefont {T.}~\bibnamefont {Esslinger}},\ }\href
  {\doibase 10.1038/nature09009} {\bibfield  {journal} {\bibinfo  {journal}
  {Nature}\ }\textbf {\bibinfo {volume} {464}},\ \bibinfo {pages} {1301}
  (\bibinfo {year} {2010})}\BibitemShut {NoStop}%
\bibitem [{\citenamefont {Stanley}(1968)}]{PhysRevLett.20.589}%
  \BibitemOpen
  \bibfield  {author} {\bibinfo {author} {\bibfnamefont {H.~E.}\ \bibnamefont
  {Stanley}},\ }\href {\doibase 10.1103/PhysRevLett.20.589} {\bibfield
  {journal} {\bibinfo  {journal} {Phys. Rev. Lett.}\ }\textbf {\bibinfo
  {volume} {20}},\ \bibinfo {pages} {589} (\bibinfo {year} {1968})}\BibitemShut
  {NoStop}%
\bibitem [{\citenamefont {Jepsen}\ \emph {et~al.}(2020)\citenamefont {Jepsen},
  \citenamefont {Amato-Grill}, \citenamefont {Dimitrova}, \citenamefont {Ho},
  \citenamefont {Demler},\ and\ \citenamefont {Ketterle}}]{RN98}%
  \BibitemOpen
  \bibfield  {author} {\bibinfo {author} {\bibfnamefont {P.~N.}\ \bibnamefont
  {Jepsen}}, \bibinfo {author} {\bibfnamefont {J.}~\bibnamefont {Amato-Grill}},
  \bibinfo {author} {\bibfnamefont {I.}~\bibnamefont {Dimitrova}}, \bibinfo
  {author} {\bibfnamefont {W.~W.}\ \bibnamefont {Ho}}, \bibinfo {author}
  {\bibfnamefont {E.}~\bibnamefont {Demler}}, \ and\ \bibinfo {author}
  {\bibfnamefont {W.}~\bibnamefont {Ketterle}},\ }\href {\doibase
  10.1038/s41586-020-3033-y} {\bibfield  {journal} {\bibinfo  {journal}
  {Nature}\ }\textbf {\bibinfo {volume} {588}},\ \bibinfo {pages} {403}
  (\bibinfo {year} {2020})}\BibitemShut {NoStop}%
\bibitem [{\citenamefont {Peierls}(1936)}]{Peierls1936}%
  \BibitemOpen
  \bibfield  {author} {\bibinfo {author} {\bibfnamefont {R.}~\bibnamefont
  {Peierls}},\ }\href {\doibase 10.1017/S0305004100019174} {\bibfield
  {journal} {\bibinfo  {journal} {Mathematical Proceedings of the Cambridge
  Philosophical Society}\ }\textbf {\bibinfo {volume} {32}},\ \bibinfo {pages}
  {477–481} (\bibinfo {year} {1936})}\BibitemShut {NoStop}%
\bibitem [{\citenamefont {Hwang}\ \emph {et~al.}(2015)\citenamefont {Hwang},
  \citenamefont {Puebla},\ and\ \citenamefont
  {Plenio}}]{PhysRevLett.115.180404}%
  \BibitemOpen
  \bibfield  {author} {\bibinfo {author} {\bibfnamefont {M.-J.}\ \bibnamefont
  {Hwang}}, \bibinfo {author} {\bibfnamefont {R.}~\bibnamefont {Puebla}}, \
  and\ \bibinfo {author} {\bibfnamefont {M.~B.}\ \bibnamefont {Plenio}},\
  }\href {\doibase 10.1103/PhysRevLett.115.180404} {\bibfield  {journal}
  {\bibinfo  {journal} {Phys. Rev. Lett.}\ }\textbf {\bibinfo {volume} {115}},\
  \bibinfo {pages} {180404} (\bibinfo {year} {2015})}\BibitemShut {NoStop}%
\bibitem [{\citenamefont {Liu}\ \emph {et~al.}(2017)\citenamefont {Liu},
  \citenamefont {Chesi}, \citenamefont {Ying}, \citenamefont {Chen},
  \citenamefont {Luo},\ and\ \citenamefont {Lin}}]{PhysRevLett.119.220601}%
  \BibitemOpen
  \bibfield  {author} {\bibinfo {author} {\bibfnamefont {M.}~\bibnamefont
  {Liu}}, \bibinfo {author} {\bibfnamefont {S.}~\bibnamefont {Chesi}}, \bibinfo
  {author} {\bibfnamefont {Z.-J.}\ \bibnamefont {Ying}}, \bibinfo {author}
  {\bibfnamefont {X.}~\bibnamefont {Chen}}, \bibinfo {author} {\bibfnamefont
  {H.-G.}\ \bibnamefont {Luo}}, \ and\ \bibinfo {author} {\bibfnamefont
  {H.-Q.}\ \bibnamefont {Lin}},\ }\href {\doibase
  10.1103/PhysRevLett.119.220601} {\bibfield  {journal} {\bibinfo  {journal}
  {Phys. Rev. Lett.}\ }\textbf {\bibinfo {volume} {119}},\ \bibinfo {pages}
  {220601} (\bibinfo {year} {2017})}\BibitemShut {NoStop}%
\bibitem [{\citenamefont {Wang}\ and\ \citenamefont
  {Hioe}(1973)}]{PhysRevA.7.831}%
  \BibitemOpen
  \bibfield  {author} {\bibinfo {author} {\bibfnamefont {Y.~K.}\ \bibnamefont
  {Wang}}\ and\ \bibinfo {author} {\bibfnamefont {F.~T.}\ \bibnamefont
  {Hioe}},\ }\href {\doibase 10.1103/PhysRevA.7.831} {\bibfield  {journal}
  {\bibinfo  {journal} {Phys. Rev. A}\ }\textbf {\bibinfo {volume} {7}},\
  \bibinfo {pages} {831} (\bibinfo {year} {1973})}\BibitemShut {NoStop}%
\bibitem [{\citenamefont {Emary}\ and\ \citenamefont
  {Brandes}(2003)}]{PhysRevE.67.066203}%
  \BibitemOpen
  \bibfield  {author} {\bibinfo {author} {\bibfnamefont {C.}~\bibnamefont
  {Emary}}\ and\ \bibinfo {author} {\bibfnamefont {T.}~\bibnamefont
  {Brandes}},\ }\href {\doibase 10.1103/PhysRevE.67.066203} {\bibfield
  {journal} {\bibinfo  {journal} {Phys. Rev. E}\ }\textbf {\bibinfo {volume}
  {67}},\ \bibinfo {pages} {066203} (\bibinfo {year} {2003})}\BibitemShut
  {NoStop}%
\bibitem [{\citenamefont {Hepp}\ and\ \citenamefont
  {Lieb}(1973)}]{HEPP1973360}%
  \BibitemOpen
  \bibfield  {author} {\bibinfo {author} {\bibfnamefont {K.}~\bibnamefont
  {Hepp}}\ and\ \bibinfo {author} {\bibfnamefont {E.~H.}\ \bibnamefont
  {Lieb}},\ }\href {\doibase https://doi.org/10.1016/0003-4916(73)90039-0}
  {\bibfield  {journal} {\bibinfo  {journal} {Ann. Phys.}\ }\textbf {\bibinfo
  {volume} {76}},\ \bibinfo {pages} {360} (\bibinfo {year} {1973})}\BibitemShut
  {NoStop}%
\bibitem [{\citenamefont {Nagy}\ \emph {et~al.}(2010)\citenamefont {Nagy},
  \citenamefont {K\'onya}, \citenamefont {Szirmai},\ and\ \citenamefont
  {Domokos}}]{PhysRevLett.104.130401}%
  \BibitemOpen
  \bibfield  {author} {\bibinfo {author} {\bibfnamefont {D.}~\bibnamefont
  {Nagy}}, \bibinfo {author} {\bibfnamefont {G.}~\bibnamefont {K\'onya}},
  \bibinfo {author} {\bibfnamefont {G.}~\bibnamefont {Szirmai}}, \ and\
  \bibinfo {author} {\bibfnamefont {P.}~\bibnamefont {Domokos}},\ }\href
  {\doibase 10.1103/PhysRevLett.104.130401} {\bibfield  {journal} {\bibinfo
  {journal} {Phys. Rev. Lett.}\ }\textbf {\bibinfo {volume} {104}},\ \bibinfo
  {pages} {130401} (\bibinfo {year} {2010})}\BibitemShut {NoStop}%
\bibitem [{\citenamefont {Baumann}\ \emph {et~al.}(2011)\citenamefont
  {Baumann}, \citenamefont {Mottl}, \citenamefont {Brennecke},\ and\
  \citenamefont {Esslinger}}]{PhysRevLett.107.140402}%
  \BibitemOpen
  \bibfield  {author} {\bibinfo {author} {\bibfnamefont {K.}~\bibnamefont
  {Baumann}}, \bibinfo {author} {\bibfnamefont {R.}~\bibnamefont {Mottl}},
  \bibinfo {author} {\bibfnamefont {F.}~\bibnamefont {Brennecke}}, \ and\
  \bibinfo {author} {\bibfnamefont {T.}~\bibnamefont {Esslinger}},\ }\href
  {\doibase 10.1103/PhysRevLett.107.140402} {\bibfield  {journal} {\bibinfo
  {journal} {Phys. Rev. Lett.}\ }\textbf {\bibinfo {volume} {107}},\ \bibinfo
  {pages} {140402} (\bibinfo {year} {2011})}\BibitemShut {NoStop}%
\bibitem [{\citenamefont {Liu}\ \emph {et~al.}(2014)\citenamefont {Liu},
  \citenamefont {Law},\ and\ \citenamefont {Ng}}]{PhysRevLett.112.086401}%
  \BibitemOpen
  \bibfield  {author} {\bibinfo {author} {\bibfnamefont {X.-J.}\ \bibnamefont
  {Liu}}, \bibinfo {author} {\bibfnamefont {K.~T.}\ \bibnamefont {Law}}, \ and\
  \bibinfo {author} {\bibfnamefont {T.~K.}\ \bibnamefont {Ng}},\ }\href
  {\doibase 10.1103/PhysRevLett.112.086401} {\bibfield  {journal} {\bibinfo
  {journal} {Phys. Rev. Lett.}\ }\textbf {\bibinfo {volume} {112}},\ \bibinfo
  {pages} {086401} (\bibinfo {year} {2014})}\BibitemShut {NoStop}%
\bibitem [{\citenamefont {Soriente}\ \emph {et~al.}(2018)\citenamefont
  {Soriente}, \citenamefont {Donner}, \citenamefont {Chitra},\ and\
  \citenamefont {Zilberberg}}]{PhysRevLett.120.183603}%
  \BibitemOpen
  \bibfield  {author} {\bibinfo {author} {\bibfnamefont {M.}~\bibnamefont
  {Soriente}}, \bibinfo {author} {\bibfnamefont {T.}~\bibnamefont {Donner}},
  \bibinfo {author} {\bibfnamefont {R.}~\bibnamefont {Chitra}}, \ and\ \bibinfo
  {author} {\bibfnamefont {O.}~\bibnamefont {Zilberberg}},\ }\href {\doibase
  10.1103/PhysRevLett.120.183603} {\bibfield  {journal} {\bibinfo  {journal}
  {Phys. Rev. Lett.}\ }\textbf {\bibinfo {volume} {120}},\ \bibinfo {pages}
  {183603} (\bibinfo {year} {2018})}\BibitemShut {NoStop}%
\bibitem [{\citenamefont {Chiacchio}\ and\ \citenamefont
  {Nunnenkamp}(2019)}]{PhysRevLett.122.193605}%
  \BibitemOpen
  \bibfield  {author} {\bibinfo {author} {\bibfnamefont {E.~I.~R.}\
  \bibnamefont {Chiacchio}}\ and\ \bibinfo {author} {\bibfnamefont
  {A.}~\bibnamefont {Nunnenkamp}},\ }\href {\doibase
  10.1103/PhysRevLett.122.193605} {\bibfield  {journal} {\bibinfo  {journal}
  {Phys. Rev. Lett.}\ }\textbf {\bibinfo {volume} {122}},\ \bibinfo {pages}
  {193605} (\bibinfo {year} {2019})}\BibitemShut {NoStop}%
\bibitem [{\citenamefont {Lin}\ \emph {et~al.}(2019)\citenamefont {Lin},
  \citenamefont {Papariello}, \citenamefont {Molignini}, \citenamefont
  {Chitra},\ and\ \citenamefont {Lode}}]{PhysRevA.100.013611}%
  \BibitemOpen
  \bibfield  {author} {\bibinfo {author} {\bibfnamefont {R.}~\bibnamefont
  {Lin}}, \bibinfo {author} {\bibfnamefont {L.}~\bibnamefont {Papariello}},
  \bibinfo {author} {\bibfnamefont {P.}~\bibnamefont {Molignini}}, \bibinfo
  {author} {\bibfnamefont {R.}~\bibnamefont {Chitra}}, \ and\ \bibinfo {author}
  {\bibfnamefont {A.~U.~J.}\ \bibnamefont {Lode}},\ }\href {\doibase
  10.1103/PhysRevA.100.013611} {\bibfield  {journal} {\bibinfo  {journal}
  {Phys. Rev. A}\ }\textbf {\bibinfo {volume} {100}},\ \bibinfo {pages}
  {013611} (\bibinfo {year} {2019})}\BibitemShut {NoStop}%
\bibitem [{\citenamefont {Yang}\ \emph {et~al.}(2021)\citenamefont {Yang},
  \citenamefont {Yin}, \citenamefont {Wen}, \citenamefont {Ji},\ and\
  \citenamefont {Sun}}]{PhysRevA.104.053313}%
  \BibitemOpen
  \bibfield  {author} {\bibinfo {author} {\bibfnamefont {M.-Y.}\ \bibnamefont
  {Yang}}, \bibinfo {author} {\bibfnamefont {H.-H.}\ \bibnamefont {Yin}},
  \bibinfo {author} {\bibfnamefont {L.}~\bibnamefont {Wen}}, \bibinfo {author}
  {\bibfnamefont {A.-C.}\ \bibnamefont {Ji}}, \ and\ \bibinfo {author}
  {\bibfnamefont {Q.}~\bibnamefont {Sun}},\ }\href {\doibase
  10.1103/PhysRevA.104.053313} {\bibfield  {journal} {\bibinfo  {journal}
  {Phys. Rev. A}\ }\textbf {\bibinfo {volume} {104}},\ \bibinfo {pages}
  {053313} (\bibinfo {year} {2021})}\BibitemShut {NoStop}%
\bibitem [{\citenamefont
  {Xu}(2026)}]{xu2026superradianceenhancessuppressesfermionic}%
  \BibitemOpen
  \bibfield  {author} {\bibinfo {author} {\bibfnamefont {Y.}~\bibnamefont
  {Xu}},\ }\href {https://arxiv.org/abs/2604.07407} {\enquote {\bibinfo {title}
  {Superradiance enhances and suppresses fermionic pairing based on universal
  critical scaling in two order parameters systems},}\ } (\bibinfo {year}
  {2026}),\ \Eprint {http://arxiv.org/abs/2604.07407} {arXiv:2604.07407
  [cond-mat.quant-gas]} \BibitemShut {NoStop}%
\bibitem [{\citenamefont {Lozano-M\'endez}\ \emph {et~al.}(2022)\citenamefont
  {Lozano-M\'endez}, \citenamefont {C\'asares},\ and\ \citenamefont
  {Caballero-Ben\'{\i}tez}}]{PhysRevLett.128.080601}%
  \BibitemOpen
  \bibfield  {author} {\bibinfo {author} {\bibfnamefont {K.}~\bibnamefont
  {Lozano-M\'endez}}, \bibinfo {author} {\bibfnamefont {A.~H.}\ \bibnamefont
  {C\'asares}}, \ and\ \bibinfo {author} {\bibfnamefont {S.~F.}\ \bibnamefont
  {Caballero-Ben\'{\i}tez}},\ }\href {\doibase 10.1103/PhysRevLett.128.080601}
  {\bibfield  {journal} {\bibinfo  {journal} {Phys. Rev. Lett.}\ }\textbf
  {\bibinfo {volume} {128}},\ \bibinfo {pages} {080601} (\bibinfo {year}
  {2022})}\BibitemShut {NoStop}%
\bibitem [{\citenamefont {Fan}\ \emph {et~al.}(2018)\citenamefont {Fan},
  \citenamefont {Zhou}, \citenamefont {Zheng}, \citenamefont {Yi},
  \citenamefont {Chen},\ and\ \citenamefont {Jia}}]{PhysRevA.98.043613}%
  \BibitemOpen
  \bibfield  {author} {\bibinfo {author} {\bibfnamefont {J.}~\bibnamefont
  {Fan}}, \bibinfo {author} {\bibfnamefont {X.}~\bibnamefont {Zhou}}, \bibinfo
  {author} {\bibfnamefont {W.}~\bibnamefont {Zheng}}, \bibinfo {author}
  {\bibfnamefont {W.}~\bibnamefont {Yi}}, \bibinfo {author} {\bibfnamefont
  {G.}~\bibnamefont {Chen}}, \ and\ \bibinfo {author} {\bibfnamefont
  {S.}~\bibnamefont {Jia}},\ }\href {\doibase 10.1103/PhysRevA.98.043613}
  {\bibfield  {journal} {\bibinfo  {journal} {Phys. Rev. A}\ }\textbf {\bibinfo
  {volume} {98}},\ \bibinfo {pages} {043613} (\bibinfo {year}
  {2018})}\BibitemShut {NoStop}%
\bibitem [{\citenamefont {Rao}\ \emph {et~al.}(2025)\citenamefont {Rao},
  \citenamefont {Lin}, \citenamefont {Luo}, \citenamefont {Guo}, \citenamefont
  {Pu},\ and\ \citenamefont
  {Gong}}]{rao2025unilateralcriticalityphasetransition}%
  \BibitemOpen
  \bibfield  {author} {\bibinfo {author} {\bibfnamefont {Z.}~\bibnamefont
  {Rao}}, \bibinfo {author} {\bibfnamefont {X.}~\bibnamefont {Lin}}, \bibinfo
  {author} {\bibfnamefont {X.}~\bibnamefont {Luo}}, \bibinfo {author}
  {\bibfnamefont {G.}~\bibnamefont {Guo}}, \bibinfo {author} {\bibfnamefont
  {H.}~\bibnamefont {Pu}}, \ and\ \bibinfo {author} {\bibfnamefont
  {M.}~\bibnamefont {Gong}},\ }\href {https://arxiv.org/abs/2509.04391}
  {\enquote {\bibinfo {title} {Unilateral criticality and phase transition in
  the cavity-ising model},}\ } (\bibinfo {year} {2025}),\ \Eprint
  {http://arxiv.org/abs/2509.04391} {arXiv:2509.04391 [quant-ph]} \BibitemShut
  {NoStop}%
\bibitem [{\citenamefont {Camacho-Guardian}\ \emph {et~al.}(2017)\citenamefont
  {Camacho-Guardian}, \citenamefont {Paredes},\ and\ \citenamefont
  {Caballero-Ben\'{\i}tez}}]{PhysRevA.96.051602}%
  \BibitemOpen
  \bibfield  {author} {\bibinfo {author} {\bibfnamefont {A.}~\bibnamefont
  {Camacho-Guardian}}, \bibinfo {author} {\bibfnamefont {R.}~\bibnamefont
  {Paredes}}, \ and\ \bibinfo {author} {\bibfnamefont {S.~F.}\ \bibnamefont
  {Caballero-Ben\'{\i}tez}},\ }\href {\doibase 10.1103/PhysRevA.96.051602}
  {\bibfield  {journal} {\bibinfo  {journal} {Phys. Rev. A}\ }\textbf {\bibinfo
  {volume} {96}},\ \bibinfo {pages} {051602(R)} (\bibinfo {year}
  {2017})}\BibitemShut {NoStop}%
\bibitem [{\citenamefont {Mivehvar}(2024)}]{PhysRevLett.132.073602}%
  \BibitemOpen
  \bibfield  {author} {\bibinfo {author} {\bibfnamefont {F.}~\bibnamefont
  {Mivehvar}},\ }\href {\doibase 10.1103/PhysRevLett.132.073602} {\bibfield
  {journal} {\bibinfo  {journal} {Phys. Rev. Lett.}\ }\textbf {\bibinfo
  {volume} {132}},\ \bibinfo {pages} {073602} (\bibinfo {year}
  {2024})}\BibitemShut {NoStop}%
\bibitem [{\citenamefont {M\"unstermann}\ \emph {et~al.}(2000)\citenamefont
  {M\"unstermann}, \citenamefont {Fischer}, \citenamefont {Maunz},
  \citenamefont {Pinkse},\ and\ \citenamefont {Rempe}}]{PhysRevLett.84.4068}%
  \BibitemOpen
  \bibfield  {author} {\bibinfo {author} {\bibfnamefont {P.}~\bibnamefont
  {M\"unstermann}}, \bibinfo {author} {\bibfnamefont {T.}~\bibnamefont
  {Fischer}}, \bibinfo {author} {\bibfnamefont {P.}~\bibnamefont {Maunz}},
  \bibinfo {author} {\bibfnamefont {P.~W.~H.}\ \bibnamefont {Pinkse}}, \ and\
  \bibinfo {author} {\bibfnamefont {G.}~\bibnamefont {Rempe}},\ }\href
  {\doibase 10.1103/PhysRevLett.84.4068} {\bibfield  {journal} {\bibinfo
  {journal} {Phys. Rev. Lett.}\ }\textbf {\bibinfo {volume} {84}},\ \bibinfo
  {pages} {4068} (\bibinfo {year} {2000})}\BibitemShut {NoStop}%
\bibitem [{\citenamefont {Gao}\ \emph {et~al.}(2020)\citenamefont {Gao},
  \citenamefont {Schlawin}, \citenamefont {Buzzi}, \citenamefont {Cavalleri},\
  and\ \citenamefont {Jaksch}}]{PhysRevLett.125.053602}%
  \BibitemOpen
  \bibfield  {author} {\bibinfo {author} {\bibfnamefont {H.}~\bibnamefont
  {Gao}}, \bibinfo {author} {\bibfnamefont {F.}~\bibnamefont {Schlawin}},
  \bibinfo {author} {\bibfnamefont {M.}~\bibnamefont {Buzzi}}, \bibinfo
  {author} {\bibfnamefont {A.}~\bibnamefont {Cavalleri}}, \ and\ \bibinfo
  {author} {\bibfnamefont {D.}~\bibnamefont {Jaksch}},\ }\href {\doibase
  10.1103/PhysRevLett.125.053602} {\bibfield  {journal} {\bibinfo  {journal}
  {Phys. Rev. Lett.}\ }\textbf {\bibinfo {volume} {125}},\ \bibinfo {pages}
  {053602} (\bibinfo {year} {2020})}\BibitemShut {NoStop}%
\bibitem [{\citenamefont {Chakraborty}\ and\ \citenamefont
  {Piazza}(2021)}]{PhysRevLett.127.177002}%
  \BibitemOpen
  \bibfield  {author} {\bibinfo {author} {\bibfnamefont {A.}~\bibnamefont
  {Chakraborty}}\ and\ \bibinfo {author} {\bibfnamefont {F.}~\bibnamefont
  {Piazza}},\ }\href {\doibase 10.1103/PhysRevLett.127.177002} {\bibfield
  {journal} {\bibinfo  {journal} {Phys. Rev. Lett.}\ }\textbf {\bibinfo
  {volume} {127}},\ \bibinfo {pages} {177002} (\bibinfo {year}
  {2021})}\BibitemShut {NoStop}%
\bibitem [{\citenamefont {Rao}\ and\ \citenamefont
  {Piazza}(2023)}]{PhysRevLett.130.083603}%
  \BibitemOpen
  \bibfield  {author} {\bibinfo {author} {\bibfnamefont {P.}~\bibnamefont
  {Rao}}\ and\ \bibinfo {author} {\bibfnamefont {F.}~\bibnamefont {Piazza}},\
  }\href {\doibase 10.1103/PhysRevLett.130.083603} {\bibfield  {journal}
  {\bibinfo  {journal} {Phys. Rev. Lett.}\ }\textbf {\bibinfo {volume} {130}},\
  \bibinfo {pages} {083603} (\bibinfo {year} {2023})}\BibitemShut {NoStop}%
\bibitem [{\citenamefont {Torre}\ \emph {et~al.}(2013)\citenamefont {Torre},
  \citenamefont {Diehl}, \citenamefont {Lukin}, \citenamefont {Sachdev},\ and\
  \citenamefont {Strack}}]{PhysRevA.87.023831}%
  \BibitemOpen
  \bibfield  {author} {\bibinfo {author} {\bibfnamefont {E.~G.~D.}\
  \bibnamefont {Torre}}, \bibinfo {author} {\bibfnamefont {S.}~\bibnamefont
  {Diehl}}, \bibinfo {author} {\bibfnamefont {M.~D.}\ \bibnamefont {Lukin}},
  \bibinfo {author} {\bibfnamefont {S.}~\bibnamefont {Sachdev}}, \ and\
  \bibinfo {author} {\bibfnamefont {P.}~\bibnamefont {Strack}},\ }\href
  {\doibase 10.1103/PhysRevA.87.023831} {\bibfield  {journal} {\bibinfo
  {journal} {Phys. Rev. A}\ }\textbf {\bibinfo {volume} {87}},\ \bibinfo
  {pages} {023831} (\bibinfo {year} {2013})}\BibitemShut {NoStop}%
\bibitem [{\citenamefont {Dmytruk}\ and\ \citenamefont
  {Schir\'o}(2021)}]{PhysRevB.103.075131}%
  \BibitemOpen
  \bibfield  {author} {\bibinfo {author} {\bibfnamefont {O.}~\bibnamefont
  {Dmytruk}}\ and\ \bibinfo {author} {\bibfnamefont {M.}~\bibnamefont
  {Schir\'o}},\ }\href {\doibase 10.1103/PhysRevB.103.075131} {\bibfield
  {journal} {\bibinfo  {journal} {Phys. Rev. B}\ }\textbf {\bibinfo {volume}
  {103}},\ \bibinfo {pages} {075131} (\bibinfo {year} {2021})}\BibitemShut
  {NoStop}%
\bibitem [{sup()}]{supplementary}%
  \BibitemOpen
  \href@noop {} {}\bibinfo {note} {See Supplemental Material for further
  information on the classification of phase diagrams and the derivation of
  critical scalings.}\BibitemShut {Stop}%
\bibitem [{\citenamefont {Xu}\ \emph {et~al.}(2024)\citenamefont {Xu},
  \citenamefont {Sun}, \citenamefont {Zhang}, \citenamefont {He},\ and\
  \citenamefont {Pu}}]{PhysRevLett.133.233604}%
  \BibitemOpen
  \bibfield  {author} {\bibinfo {author} {\bibfnamefont {Y.}~\bibnamefont
  {Xu}}, \bibinfo {author} {\bibfnamefont {F.-X.}\ \bibnamefont {Sun}},
  \bibinfo {author} {\bibfnamefont {W.}~\bibnamefont {Zhang}}, \bibinfo
  {author} {\bibfnamefont {Q.}~\bibnamefont {He}}, \ and\ \bibinfo {author}
  {\bibfnamefont {H.}~\bibnamefont {Pu}},\ }\href {\doibase
  10.1103/PhysRevLett.133.233604} {\bibfield  {journal} {\bibinfo  {journal}
  {Phys. Rev. Lett.}\ }\textbf {\bibinfo {volume} {133}},\ \bibinfo {pages}
  {233604} (\bibinfo {year} {2024})}\BibitemShut {NoStop}%
\bibitem [{\citenamefont {Baksic}\ and\ \citenamefont
  {Ciuti}(2014)}]{PhysRevLett.112.173601}%
  \BibitemOpen
  \bibfield  {author} {\bibinfo {author} {\bibfnamefont {A.}~\bibnamefont
  {Baksic}}\ and\ \bibinfo {author} {\bibfnamefont {C.}~\bibnamefont {Ciuti}},\
  }\href {\doibase 10.1103/PhysRevLett.112.173601} {\bibfield  {journal}
  {\bibinfo  {journal} {Phys. Rev. Lett.}\ }\textbf {\bibinfo {volume} {112}},\
  \bibinfo {pages} {173601} (\bibinfo {year} {2014})}\BibitemShut {NoStop}%
\end{thebibliography}%
		
		\onecolumngrid
		
		\section{universal critical scaling rate of continuous phase transtion in two order parameters coexisted system}
		
		In Landau's paradigm of phase transition, the macroscopic order parameters, such as photon number, magnetism, disorder, can be viewed as the variables in the physical systems, and the free energy can be viewed as the function against these variables. By minimizing the free energy of the system, one can obtain the order parameters of the ground states or the stable states.
		
		Consider a system described by two macroscopic order parameters $\left(O_1,O_2\right)$ with the environmental parameter vector $\vec{\lambda}$, giving as
		\begin{align}\label{free_energy}
			F=F(O_1,O_2;\vec{\lambda})=\sum_{v_1,v_2}g_{v_1,v_2}(\vec{\lambda})O_1^{v_1}O_2^{v_2}.
		\end{align}
		Generally, the order parameter $O_i$ is well selected as nonzero real numbers, such as the number of photon in the superradiant phase transition or the absolute value of the mean magnetism in the ferromagnetic phase. And the system status will be determined by the order parameters $\left(\bar{O}_1(\vec{\lambda}),\bar{O}_2(\vec{\lambda})\right)$, where the free energy is minimum, i.e., it satisfies
		\begin{align}
			F(\bar{O}_1(\vec{\lambda}),\bar{O}_2(\vec{\lambda}))=\min_{(O_1,O_2)}F(O_1,O_2;\vec{\lambda}).
		\end{align}
		Notice that the system status $\left(\bar{O}_1(\vec{\lambda}),\bar{O}_2(\vec{\lambda})\right)$ are the functions against the environmental parameter $\vec{\lambda}$. 
		The phases with order parameter $\bar{O}_i=0$ and $\bar{O}_i\neq0$ represent the normal phase and the ordered phase respectively, which are seperated by the critical point $\vec{\lambda}_c$. At the minimum point, we have $\dfrac{\partial F(\vec{\lambda})}{\partial O_i}\Bigg|_{\left(\bar{O}_1(\vec{\lambda}),\bar{O}_2(\vec{\lambda})\right)}\equiv A_i\left(\bar{O}_1(\vec{\lambda}),\bar{O}_2(\vec{\lambda});\vec{\lambda}\right)=0$. Also, the Hessian matrix is positive, i.e.,  $\det[\mathcal{H}(\vec{\lambda})\Bigg|_{\left(\bar{O}_1(\vec{\lambda}),\bar{O}_2(\vec{\lambda})\right)}]>0$, where the Hessian matrix is defined based on the 2nd order differentiations of the free energy, reads
		\begin{align}
			\mathcal{H}(\vec{\lambda})\Bigg|_{\left(\bar{O}_1(\vec{\lambda}),\bar{O}_2(\vec{\lambda})\right)}\equiv\left(\begin{array}{cc}
				\dfrac{\partial^2}{\partial O_1^2}&\dfrac{\partial^2}{\partial O_1\partial O_2}\\
				\dfrac{\partial^2}{\partial O_1\partial O_2}&\dfrac{\partial^2}{\partial O_2^2}
			\end{array}\right)F(\vec{\lambda})\Bigg|_{\left(\bar{O}_1(\vec{\lambda}),\bar{O}_2(\vec{\lambda})\right)}\equiv\left(\begin{array}{cc}
				C_{11}\left(\bar{O}_1(\vec{\lambda}),\bar{O}_2(\vec{\lambda});\vec{\lambda}\right)&C_{12}\left(\bar{O}_1(\vec{\lambda}),\bar{O}_2(\vec{\lambda});\vec{\lambda}\right)\\
				C_{21}\left(\bar{O}_1(\vec{\lambda}),\bar{O}_2(\vec{\lambda});\vec{\lambda}\right)&C_{22}\left(\bar{O}_1(\vec{\lambda}),\bar{O}_2(\vec{\lambda});\vec{\lambda}\right)
			\end{array}\right).
		\end{align}
		
		Given that the general free energy~(\ref{free_energy}) is rather complex, we give some assumptions before the following discussions. On one hand, we consider the linear term is absent, where $g_{1,0}(\vec{\lambda})=g_{0,1}(\vec{\lambda})=0,~\forall\vec{\lambda}$, in order to ganrantee the existence of normal phase $\bar{O}_i=0$. On the other hand, we assume the lowest order terms is quadratic. In other words, the situation $g_{2,0}(\vec{\lambda})g_{0,2}(\vec{\lambda})=0,~\forall\vec{\lambda}$ is excluded. Additionally, we only concentrate on the continuous phase transition, whose critical points and scaling rates are analytically accessible. 
		
		
		Without loss of generality, we consider the 2D parametric plane to demonstrate the universal phase diagrams, where $\vec{\lambda}=(\lambda_1,\lambda_2)$. 
		The general $d-$dimensional space spanned by vector $\vec{\lambda}$ space can be demonstrated similarly, where we also keep the 2 dimensions $\lambda_{i_1},\lambda_{i_2}$ variable and the other components $j\neq i_1,i_2$ fixed. Then, we will categorize all the phase diagrams into four different categories according to the coupling terms in Eq.~(\ref{free_energy}).
		
		\subsection{The analytical phase boundaries}
		
		
		As is shown in Fig. 1 of the main text, the typical patterns are listed in all categories. Before detailed evaluations on each pattern and the critical boundary, we first give several theorems as follows to determine the critical conditions for continous phase transition.
		
		(i)Firstly, we give a simple theorem on the continously phase transition for single order parameter $O_1$ with $\bar{O}_2=0$. If system status is given by $\left(\bar{O}_1=0,\bar{O}_2=0\right)$ before the continous critical boundary $l$, crossing which it changes into $\left(\bar{O}_1\neq0,\bar{O}_2=0\right)$. Then the critical boundary is expressed as $C_{11}(0,0;\vec{\lambda}_c)=0$, with $C_{22}(0,0;\vec{\lambda}_c)\ge0$.
		
		Proof: Consider that the $\left(0,0\right)$ is still stable on the continous critical line, so $C_{11}(0,0;\vec{\lambda}_c)<0$ is excluded. Assume that $C_{11}(0,0;\vec{\lambda}_c)>0$ on the line, then we can always select a small $\left|\delta\vec{\lambda}\right|$ making $C_{11}(0,0;\vec{\lambda}_c+\delta\vec{\lambda})>0$ with $\vec{\lambda}_c+\delta\vec{\lambda}$ across the $l$. Consider all the linear term is absent, we have
		\begin{align}
			A_1(0,0;\vec{\lambda}_c+\delta\vec{\lambda})=0,~~C_{11}(0,0;\vec{\lambda}_c+\delta\vec{\lambda})>0.
		\end{align}
		Then we consider the stability of the new configuration $\left(\bar{O}_1(\vec{\lambda}_c+\delta\vec{\lambda})\neq0,0\right)$ is supported, satisfying 
		\begin{align}
			A_1\left(\bar{O}_1(\vec{\lambda}_c+\delta\vec{\lambda}),0;\vec{\lambda}_c+\delta\vec{\lambda}\right)=0,~~C_{11}\left(\bar{O}_1(\vec{\lambda}_c+\delta\vec{\lambda}),0;\vec{\lambda}_c+\delta\vec{\lambda}\right)>0.
		\end{align}
		Compare the above two conditions, we can deduce that the $\bar{O}_1(\vec{\lambda}_c+\delta\vec{\lambda})$ is finite, despite $\left|\delta\vec{\lambda}\right|\to0$. So it contradicts the continuity of the boundary. So the only $C_{11}(0,0;\vec{\lambda}_c)=0$ can describe the critical line $l$, and $C_{22}(0,0;\vec{\lambda}_c)\ge0$ keeps the stability in the direction $O_2$.
		
		(ii) The conclusion can be extended to a general case where the critical line continuously splits two region $\left(\bar{O}_1=0,\bar{O}_2\right)$ and $\left(\bar{O}_1\neq0,\bar{O}_2\right)$ for arbitray $\bar{O}_2$. We emphasize that the $\bar{O}_2$ also keeps continuity on the boundary. And the linear condition is changed into $g_{1,v_2}(\vec{\lambda}_c)=0$ for arbitrary $v_2$, indicating $A_1(0,\bar{O}_2;\vec{\lambda}_c)=\sum_{v_2}g_{1,v_2}(\vec{\lambda}_c)\bar{O}_2^{v_2}=0$. In this case, the critical boundary is given as $C_{11}\left(0,\bar{O}_2(\vec{\lambda}_c);\vec{\lambda}_c\right)=0$.
		
		Proof: Consider that the $\left(\bar{O}_1=0,\bar{O}_2(\vec{\lambda}_c)\right)$ is still stable on the continous critical line, so $C_{11}(0,\bar{O}_2(\vec{\lambda}_c);\vec{\lambda}_c)<0$ is excluded. Assume that $C_{11}(0,\bar{O}_2(\vec{\lambda}_c);\vec{\lambda}_c)>0$ on the line, then we can always select a small $\left|\delta\vec{\lambda}\right|$ making $C_{11}(0,\bar{O}_2(\vec{\lambda}_c+\delta\vec{\lambda});\vec{\lambda}_c+\delta\vec{\lambda})>0$ with $\vec{\lambda}_c+\delta\vec{\lambda}$ across the $l$. Consider the linearity condition, we have
		\begin{align}
			A_1(0,\bar{O}_2(\vec{\lambda}_c+\delta\vec{\lambda});\vec{\lambda}_c+\delta\vec{\lambda})=0,~~C_{11}(0,\bar{O}_2(\vec{\lambda}_c+\delta\vec{\lambda});\vec{\lambda}_c+\delta\vec{\lambda})>0.
		\end{align}
		Then we consider the stability of the new configuration $\left(\bar{O}_1(\vec{\lambda}_c+\delta\vec{\lambda})\neq0,\bar{O}_2(\vec{\lambda}_c+\delta\vec{\lambda})\right)$ is supported, satisfying 
		\begin{align}
			A_1\left(\bar{O}_1(\vec{\lambda}_c+\delta\vec{\lambda}),\bar{O}_2(\vec{\lambda}_c+\delta\vec{\lambda});\vec{\lambda}_c+\delta\vec{\lambda}\right)=0,~~C_{11}\left(\bar{O}_1(\vec{\lambda}_c+\delta\vec{\lambda}),\bar{O}_2(\vec{\lambda}_c+\delta\vec{\lambda});\vec{\lambda}_c+\delta\vec{\lambda}\right)>0.
		\end{align}
		Compare the above two conditions, we can deduce that the $\bar{O}_1(\vec{\lambda}_c+\delta\vec{\lambda})$ is finite, despite $\left|\delta\vec{\lambda}\right|\to0$. So it contradicts the continuity of the boundary. So the only $C_{11}(0,\bar{O}_2(\vec{\lambda}_c);\vec{\lambda}_c)=0$ can describe the critical line $l$.
		
		(iii) Meanwhile, the conclusion (i) can be generalized to higher order case. Take two order parameter case for instance, if the continous phase boundary splits two phases with $\left(\bar{O}_1=0,\bar{O}_2=0\right)$ and $\left(\bar{O}_1\neq0,\bar{O}_2\neq0\right)$, then the critical boundary is given as the Hessian matrix $\det[\mathcal{H}(\vec{\lambda}_c)\Bigg|_{0,0}]=0$, with the Hessian matrix positive semi-definite according to the stability.
		
		proof: Consider that the point $\left(\bar{O}_1=0,\bar{O}_2=0\right)$ is stable on the critical line, so the Hessian matrix must be positive semi-definite with $\mathcal{H}(\vec{\lambda}_c)\Bigg|_{0,0}\ge0$. Suppose that the Hessian matrix is positive definite, i.e., $\mathcal{H}(\vec{\lambda}_c)\Bigg|_{0,0}>0$ on the critical line, then we can always select a small $\left|\delta\vec{\lambda}\right|$ making $\mathcal{H}(\vec{\lambda}_c+\delta\vec{\lambda})\Bigg|_{0,0}>0$ with $\vec{\lambda}_c+\delta\vec{\lambda}$ across the $l$. Consider the linear term is absent, we have
		\begin{align}
			(\vec{n}\cdot\nabla)F(\vec{\lambda}_c+\delta\vec{\lambda})\Bigg|_{(0,0)}=0,~~(\vec{n}\cdot\nabla)^2F(\vec{\lambda}_c+\delta\vec{\lambda})\Bigg|_{(0,0)}>0
		\end{align}
		for arbitrary direction $\vec{n}$ in the space spanned by $O_1,~O_2$. Then we consider the stability of the new configuration $\left(\bar{O}_1(\vec{\lambda}_c+\delta\vec{\lambda})\neq0,\bar{O}_2(\vec{\lambda}_c+\delta\vec{\lambda})\neq0\right)$ is supported, gives
		\begin{align}
			(\vec{n}\cdot\nabla)F(\vec{\lambda}_c+\delta\vec{\lambda})\Bigg|_{\left(\bar{O}_1(\vec{\lambda}_c+\delta\vec{\lambda}),\bar{O}_2(\vec{\lambda}_c+\delta\vec{\lambda})\right)}=0,~~(\vec{n}\cdot\nabla)^2F(\vec{\lambda}_c+\delta\vec{\lambda})\Bigg|_{\left(\bar{O}_1(\vec{\lambda}_c+\delta\vec{\lambda}),\bar{O}_2(\vec{\lambda}_c+\delta\vec{\lambda})\right)}>0
		\end{align} 
		for arbitrary direction $\vec{n}$ in the space spanned by $O_1,~O_2$. Choose the direction as $\vec{n}=\dfrac{\bar{O}_1(\vec{\lambda}_c+\delta\vec{\lambda})\vec{e}_1+\bar{O}_2(\vec{\lambda}_c+\delta\vec{\lambda})\vec{e}_2}{\sqrt{\bar{O}_1^2(\vec{\lambda}_c+\delta\vec{\lambda})+\bar{O}_2^2(\vec{\lambda}_c+\delta\vec{\lambda})}}$, this situation is reduced to the 1 dimensional case, where we can see the positive definite Hessian matrix contradicts the continuity of the phase transition. So we can get the only correct description of the critical line as $\det[\mathcal{H}(\vec{\lambda}_c)\Bigg|_{0,0}]=0$, implying at least one eigenvalue of the Hessian matrix is zero with the Hessian matrix positive semi-definite simultanuously. 
		
		(iv) The conclusions (i-iii) can be generalized to higher dimensional cases in multi-order parameters system following the similar approach.
		
		
		\begin{figure}[tb]\label{patterns}
			\centering
			\includegraphics[width=0.48\textwidth]{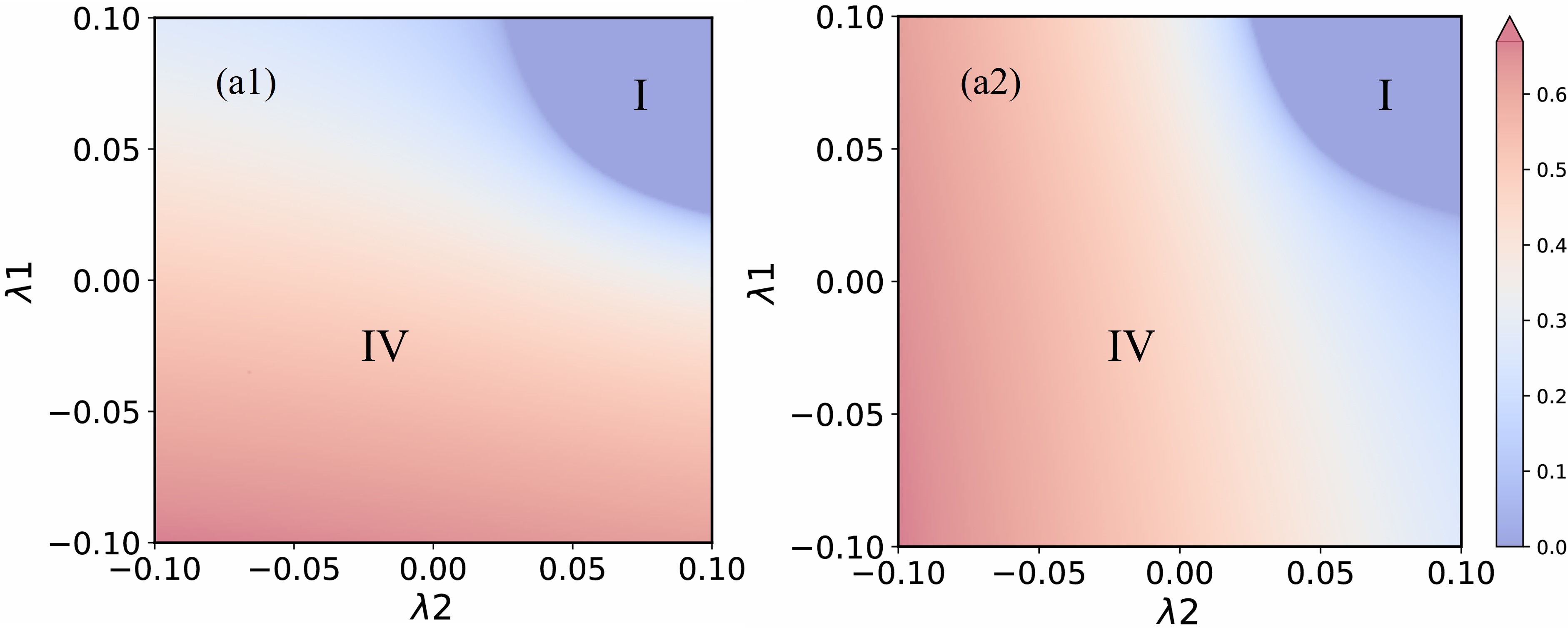}
			\includegraphics[width=0.48\textwidth]{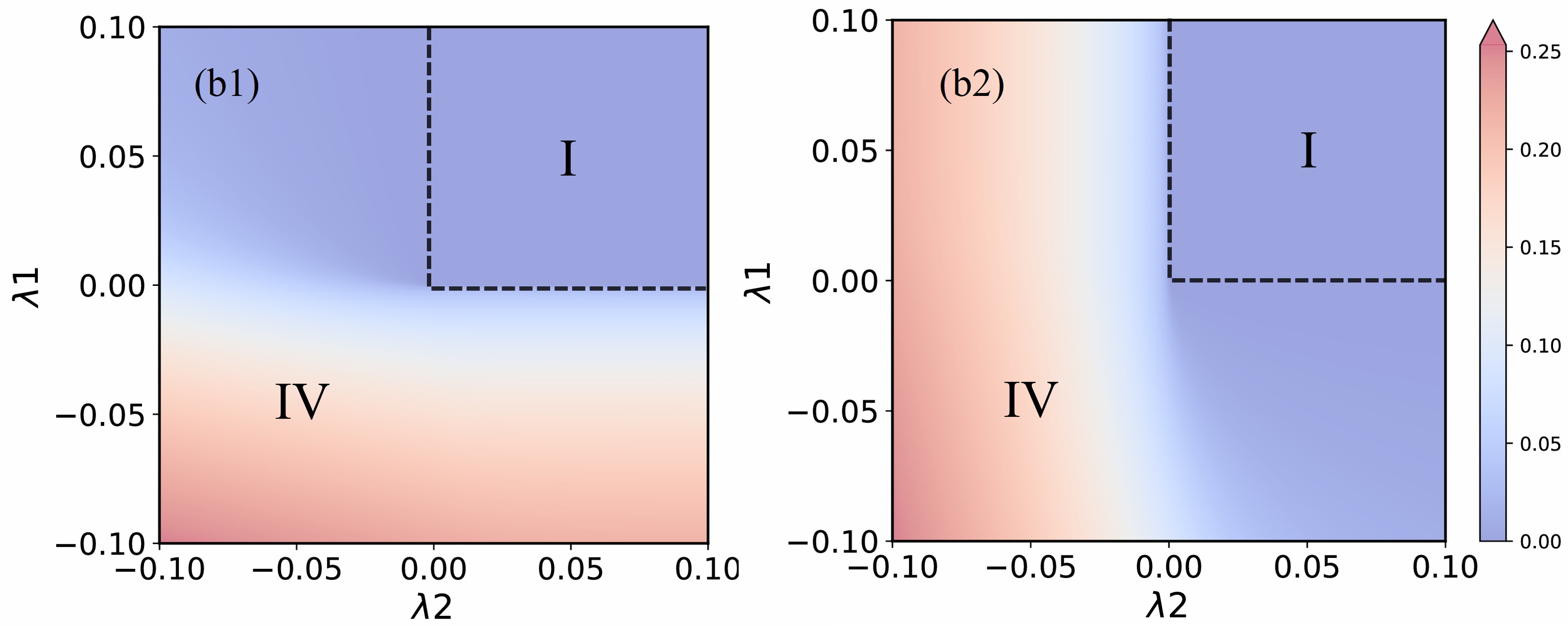}
			\includegraphics[width=0.48\textwidth]{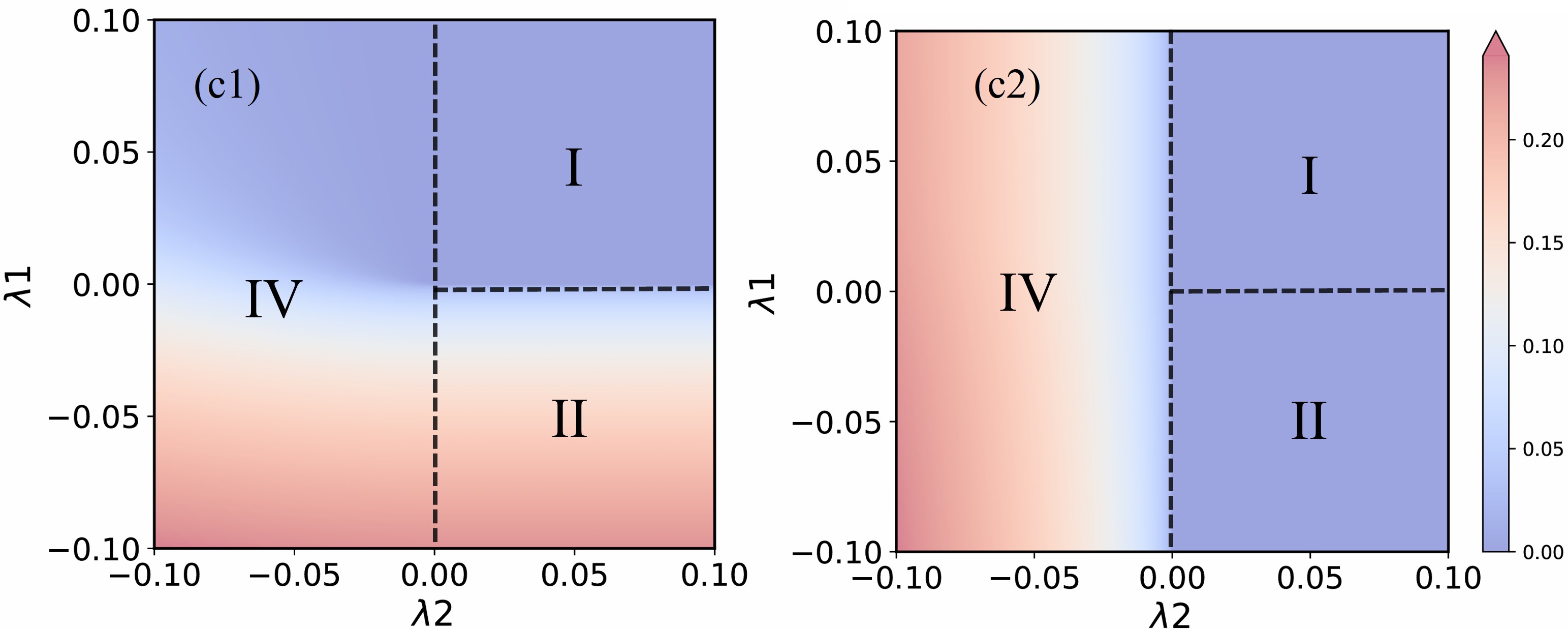}
			\includegraphics[width=0.48\textwidth]{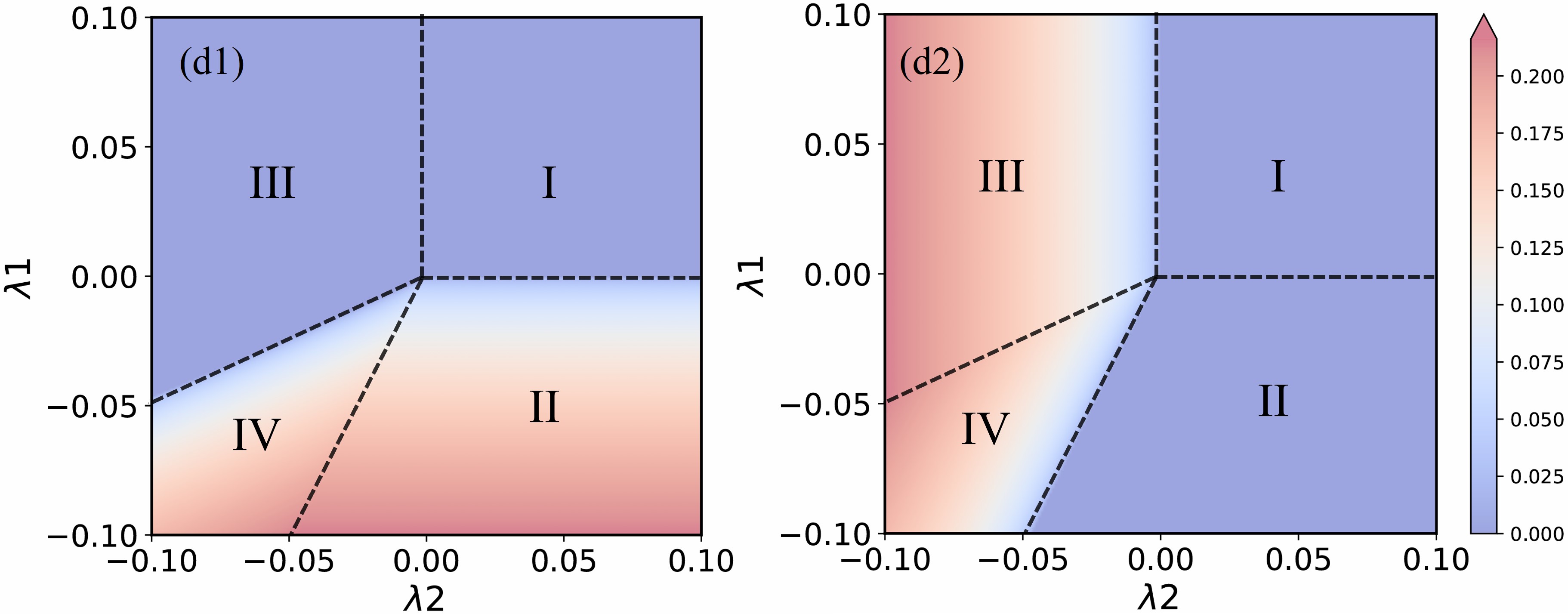}
			\caption{The typical patterns of phase diagram for Category A, $A_2$, B and C are shown in (a1)(a2), (b1)(b2), (c1)(c2) and (d1)(d2) respectively.}
		\end{figure}
		
		\subsection{Category A}
		
		In category A, where the coupling term satisfies $g_{1,1}(\vec{\lambda})\neq0$, the typical configurations are $\left(\bar{O}_1(\vec{\lambda})=0,\bar{O}_2(\vec{\lambda})=0\right)$ and $\left(\bar{O}_1(\vec{\lambda})\neq0,\bar{O}_2(\vec{\lambda})\neq0\right)$ denoted by type-I and IV respectively. And the other two types of minimum points $\left(\bar{O}_1(\vec{\lambda})\neq0,\bar{O}_2(\vec{\lambda})=0\right)$ and $\left(\bar{O}_1(\vec{\lambda})=0,\bar{O}_2(\vec{\lambda})\neq0\right)$, i.e., type-II and III configurations are excluded, which can be checked by $A_2(\bar{O}_1\neq0,0;\vec{\lambda})=\sum_{v_1}g_{v_1,1}(\vec{\lambda})\bar{O}_1^{v_1}$ and $A_1(0,\bar{O}_2\neq0;\vec{\lambda})=\sum_{v_2}g_{1,v_2}(\vec{\lambda})\bar{O}_2^{v_2}$, which hardly become zero unless at the well selected parametric value $\vec{\lambda}$, and they're not the "phase" that can characterize the common feature of the system in a stable and well-defined parametric region. So such status won't emergy as the "phase" here.
		
		The typical continuous phase transition is only the one from type-I to type-IV, which is seperated by the critical boundary expressed as 
		\begin{align}\label{critical_boundary_A}
			l:~C_{11}(0,0;\vec{\lambda}_c)C_{22}(0,0;\vec{\lambda}_c)-C_{12}^2(0,0;\vec{\lambda}_c)=0
		\end{align}
		in $C_{11}(0,0;\vec{\lambda}_c),C_{22}(0,0;\vec{\lambda}_c)\ge0$ brunch according to the theorem (iii).
		
		\subsection{Category $A_2$}
		
		In category $A_2$, the coupling $g_{1,1}(\vec{\lambda})=0$, but $g_{v_1,1}(\vec{\lambda})\neq0,~\exists v_1$ and $g_{1,v_2}(\vec{\lambda})\neq0,~\exists v_2$ are both satisfied. In this case, the type-II and III are both absent due to $A_1(0,\bar{O}_2\neq0;\vec{\lambda})=\sum_{v_2}g_{1,v_2}(\vec{\lambda})\bar{O}_2^{v_2}$ and $A_2(\bar{O}_1\neq0,0;\vec{\lambda})=\sum_{v_1}g_{v_1,1}(\vec{\lambda})\bar{O}_1^{v_1}$. For the same reason as above, they can't be viewed as phases here. Meanwhile, distinguished from the Category A, $C_{12}(0,0;\vec{\lambda})=0$ is satisfied in Category $A_2$. So, the critical boundary can be expressed as $C_{11}(0,0;\vec{\lambda}_c)=0$ with $C_{22}(0,0;\vec{\lambda}_c)\ge0$, and $C_{22}(0,0;\vec{\lambda}_c)=0$ with $C_{11}(0,0;\vec{\lambda}_c)\ge0$. The two independent curves consists of the real boundary between type-I and IV.
		
		Although the Category A and $A_2$ both supports only the type-I and IV phases, both the critical boundary and the critical behaviour are absolutely different, which will be discussed later. 
		
		\subsection{Category B}
		For Category B, we consider $g_{v_1,1}(\vec{\lambda})=0,~\forall v_1$, i.e., the terms like $O_1^{v_1}O_2$ are excluded, while $g_{1,v_2}(\vec{\lambda})\neq0,~\exists v_2$ is satisfied, i.e., the terms like $O_1O_2^{v_2}$ are taken into consideration. In this case, type-I, II and IV configurations may be supported, but the type-III is absent by calculation as $A_1(0,\bar{O}_2\neq0;\vec{\lambda})=\sum_{v_2}g_{1,v_2}(\vec{\lambda})\bar{O}_2^{v_2}$. For the same reason as before, it can't be viewed a "phase" here.

		The general continuous phase transtion is divided into three parts in this case.
		
		(1)Type-I $\to$ II: This transition keeps the $\bar{O}_2=0$ and gives a continuous transition $\bar{O}_1=0\to\bar{O}_1\neq0$. According to the theorem (i), the critical boundary can write as $l_{12}:~~C_{11}(0,0;\vec{\lambda}_c)=0$, with $C_{22}(0,0;\vec{\lambda}_c)\ge0$.
		
		(2)Type-II $\to$ IV: This transition keeps the continuity of $\bar{O}_1$, but gives the transition $\bar{O}_2=0\to\bar{O}_2\neq0$. According to the theorem (ii), the critical boundary can write as $l_{24}:~~C_{22}(\bar{O}_1,0;\vec{\lambda}_c)=0$.
		
		(3)Type-I $\to$ IV: This transition gives the change from $\left(\bar{O}_1=0,\bar{O}_2=0\right)$ to $\left(\bar{O}_1\neq0,\bar{O}_2\neq0\right)$. It seems that we can identify the critical boundary as $\det[\mathcal{H}(\vec{\lambda}_c)\Bigg|_{\left(0,0\right)}]=0$. But it turns out that the $C_{12}(0,0;\vec{\lambda})=0$, giving us two independent boundary as $C_{11}(0,0;\vec{\lambda}_c)=0$ with $C_{22}(0,0;\vec{\lambda}_c)\ge0$, and $C_{22}(0,0;\vec{\lambda}_c)=0$ with $C_{11}(0,0;\vec{\lambda}_c)\ge0$. We find the former one is nothing but $l_{12}$. If the critical boundary is also expressed by $l_{12}$, the transition $\bar{O}_2=0\to\bar{O}_2\neq0$ in region $C_{22}(0,0;\vec{\lambda}_c)>0$ will be discontinuous by means of the same procedure in the proof of theorem (ii). Then it will contradict the continuity of the transition. So the critical boundary is $l_{14}:~~C_{22}(0,0;\vec{\lambda}_c)=0$ with $C_{11}(0,0;\vec{\lambda}_c)\ge0$. [Actually, we can directly apply the theorem (ii).]
		
		Distinguished from the Category A, the tricritical point may occur at $P_B:~~C_{11}(0,0;\vec{\lambda}_t)=C_{22}(0,0;\vec{\lambda}_t)=0$.

		\subsection{Category C}
		In category C, we consider $g_{v_1,1}(\vec{\lambda})=g_{1,v_2}(\vec{\lambda})=0,~\forall v_1,v_2$. In this case, the four types of configurations are all supported in this type. 
		
		(1)Type-I $\to$ II: This transition keeps the $\bar{O}_2=0$ and gives a continuous transition $\bar{O}_1=0\to\bar{O}_1\neq0$. The critical boundary is writen as $l_{12}:~~C_{11}(0,0;\vec{\lambda}_c)=0$, with $C_{22}(0,0;\vec{\lambda}_c)\ge0$.
		
		(2)Type-II $\to$ IV: This transition keeps the continuity of $\bar{O}_1$, but gives the transition $\bar{O}_2=0\to\bar{O}_2\neq0$. The critical boundary is given as $l_{24}:~~C_{22}(\bar{O}_1,0;\vec{\lambda}_c)=0$.
		
		(3)Type-I $\to$ III: This transition keeps the $\bar{O}_1=0$ and gives a continuous transition $\bar{O}_2=0\to\bar{O}_2\neq0$. As the counterpart of the transition from I $\to$ III, the critical boundary can write as $l_{13}:~~C_{22}(0,0;\vec{\lambda}_c)=0$, with $C_{11}(0,0;\vec{\lambda}_c)\ge0$.
		
		(4)Type-III $\to$ IV: This transition keeps the continuity of $\bar{O}_2$, but gives the transition $\bar{O}_1=0\to\bar{O}_1\neq0$. As the counterpart of the transition from II to IV, the critical boundary can write as $l_{34}:~~C_{11}(0,\bar{O}_2;\vec{\lambda}_c)=0$.
		
		Notice that, similar to the analysis (3) in the last section, or directly apply the theorem (ii), we can see that the critical point between type-I and type-IV must on both $l_{12}$ and $l_{13}$, i.e., the intersection between $C_{11}(0,0;\vec{\lambda}_c)=0$ and $C_{22}(0,0;\vec{\lambda}_c)=0$. Also the continuous transition from type-II and type-III must occur from $\bar{O}_1\neq0$ to $\bar{O}_1=0$ and from $\bar{O}_2=0$ to $\bar{O}_2\neq0$, thus the critical point is also located on the intersection between $C_{11}(0,0;\vec{\lambda}_c)=0$ and $C_{22}(0,0;\vec{\lambda}_c)=0$. Therefore, distinguished from the Category A and B, the multicritical point may occur at $P_B:~~C_{11}(0,0;\vec{\lambda}_m)=C_{22}(0,0;\vec{\lambda}_m)=0$.

		\subsection{The linear case}
		In above discussion, we asume that the linear terms are absent to guarantee the existence of phase I. For a more general case where one linear term exists, we can view it as the special case for the above three categories. Consider $g_{1,0}(\vec{\lambda})$ is nonzero generally, while $g_{v_1,1}(\vec{\lambda})=0,~\forall v_1$. The stable status $\left(0,\bar{O}_2\right)$ is excluded by checking the 1st order differentiation as $A_1(0,\bar{O}_2;\vec{\lambda})=\sum_{v_2}g_{1,v_2}(\vec{\lambda})\bar{O}_2^{v_2}=g_{1,0}(\vec{\lambda})+\sum_{v_2\neq0}g_{1,v_2}(\vec{\lambda})\bar{O}_2^{v_2}$, generally giving a nonzero value unless a well selected parametric point $\vec{\lambda}$, which can't be viewed as a "phase". Therefore, only the type-II and IV configuration is supported. According to the theorem (ii), the continuous phase boundary is given as $l_{24}:~~C_{22}(\bar{O}_1,0;\vec{\lambda}_c)=0$.
		
		We can easily check that if $g_{1,0}(\vec{\lambda})\neq0$ generally and $g_{v_1,1}(\vec{\lambda})\neq0,~\exists v_1$, i.e., the linear terms exist in both directions of order parameter, only the configuration type-IV will survive here, indicating no continuous phase transition.
		
		\subsection{The numerical phase diagrams based on minimization of free energy}
		
		In order for explicit illustration, we give the phase diagrams of the category A, $A_2$, B and C in Fig.~\ref{patterns} as four examples by numerically solving the minimum points of the self-defined free energy $F_A$, $F_{A_2}$, $F_B$ and $F_C$ respectively. We set the expressions of the free energy as
		\begin{align}
			F_A&=\lambda_1O_1^2+\lambda_2O_2^2+0.1O_1O_2+0.1(O_1^4+O_2^4)+0.1(O_1^6+O_2^6)\\
			F_{A_2}&=\lambda_1O_1^2+\lambda_2O_2^2-O_1O_2^4-O_1^4O_2+O_1^4+O_2^4+O_1^6+O_2^6\\
			F_B&=\lambda_1O_1^2+\lambda_2O_2^2-O_1O_2^4+O_1^4+O_2^4+O_1^6+O_2^6\\
			F_C&=\lambda_1O_1^2+\lambda_2O_2^2+O_1^2O_2^2+O_1^4+O_2^4+O_1^6+O_2^6.
		\end{align}
		Here, we set the coefficients before the quadratic terms as the environmental parameters $\lambda_1$, $\lambda_2$. 
		
		According to our classification, the free energy $F_A$ gives the coupling term $0.1O_1O_2$, corresponding to Category A. The subfigure (a1) and (a2) demonstrate the colormaps of order paramter $\bar{O}_1$ and $\bar{O}_2$ respectively, and the critical boundary is given by $\lambda_1\lambda_2=0.05^2$ with $\lambda_{1(2)}\ge0$. 
		
		Then the free energy $F_{A_2}$ contains the terms $O_1O_2^4$ and $O_1^4O_2$ without the quadratic coupling term $O_1O_2$. It refers to the Category $A_2$ in our classification, giving the phase boundary $\lambda_{1(2)}=0$ with $\lambda_{2(1)}\ge0$ shown in subfigure (b1) and (b2).
		
		Similarly, consider the expression $F_B$, where only the term $O_1O_2^4$ involved, corresponding to the Category B. The transition from phase I to II(IV) is $\lambda_{1(2)}=0$ with $\lambda_{2(1)}\ge0$, while the boundary between II to IV is given by $\dfrac{\partial^2 F_B}{2\partial O_2^2}\Bigg|_{O_1=\bar{O}_1,O_2=0}=\lambda_2=0$ with $\lambda_1<0$. The phase diagrams are shown in subfigure (c1) and (c2).
		
		Lastly, the free energy expression $F_C$ is classified as the Category C, The transition from phase I to II(III) is $\lambda_{1(2)}=0$ with $\lambda_{2(1)}\ge0$
		$(O_1^2)^3+(O_1^2)^2+\lambda_1O_1^2$. The critical boundary between phase II and IV is obtained as $\dfrac{\partial^2 F_B}{2\partial O_2^2}\Bigg|_{O_1=\bar{O}_1,O_2=0}=\lambda_2+\bar{O}_1^2=0$, and the $\bar{O}_1$ is the minimum position as $O_2=0$, obtained by the solution of the following equation
		\begin{align}
			\dfrac{\partial F_C}{\partial(O_1^2)}\Bigg|_{O_2=0}=3(O_1^2)^2+2O_1^2+\lambda_1=0.
		\end{align}
		
		It gives the solution $\bar{O}_1^2=\dfrac{-1+\sqrt{1-3\lambda_1}}{3}$. Inserting it into the critical condition, gives the boundary $3\lambda_2+\sqrt{1-3\lambda_1}=1$ with $\lambda_1<0$. Similarly, the critical boundary between phase III and IV is expressed as $3\lambda_1+\sqrt{1-3\lambda_2}=1$ with $\lambda_2<0$. The phase diagrams of free energy $F_C$ are shown in subfigure (d1) and (d2).
		
		{\color{blue}\section{The real structure of phase diagram}
			
			As is mentioned in the main text, the real situations are much more complex than the typical patterns displayed in Fig.1. Because the value ranges of the function cluster $\left\{g_{v_1,v_2}(\vec{\lambda})\right\}$ decide the real structure of phase diagram. Take the Category B as the example, three different kinds of real structure are named as the fragment, defections and splice are detailed illustrated in the first, second and thirs rows in Fig.~\ref{supp_real} respectively.
			
			The fragments are a serie of simple phase diagrams with at most three phases I, II and IV (in Category B), and each phase occupies an individual connected region. All the situation are further divided as the first row according to the existence of tricritical point and the critical boundary. The case (a1) with one tricritical point is nothing but the typical pattern in the Table. The cases (a2) are the fragments with only one critical boundaries seperating two phases, and the tricritical point is absent. The fragments with only one phase without either the tricritical point or the critical boundary are shown in (a3).
			
			The second row gives the categories of the defection, corresponding to the first row one by one. The defections are originated from the 1st order phase transition, marked by the shadows in the diagrams. (b1) shows the case where the defection covers the tricritical point in (a1), while (b2) gives the situations that the critical boundaries are affected. The last situation is the defections inside the phases shown in (b3).
			
			The splices are viewed as the constructions of several sub-blocks. The sub-block can be the fragments, the defections or other splices. Some examples are listed in the last row. The splices combined by two fragments (a2), the fragment (a1) with the defection (b2), and the fragment (a1) with the defection (b1) are arranged from left to right. The sub-blocks are divided by the dotted lines.
			
			\begin{figure}[tb]
				\centering
				\hspace{-0.6cm}
				\includegraphics[width=0.96\textwidth]{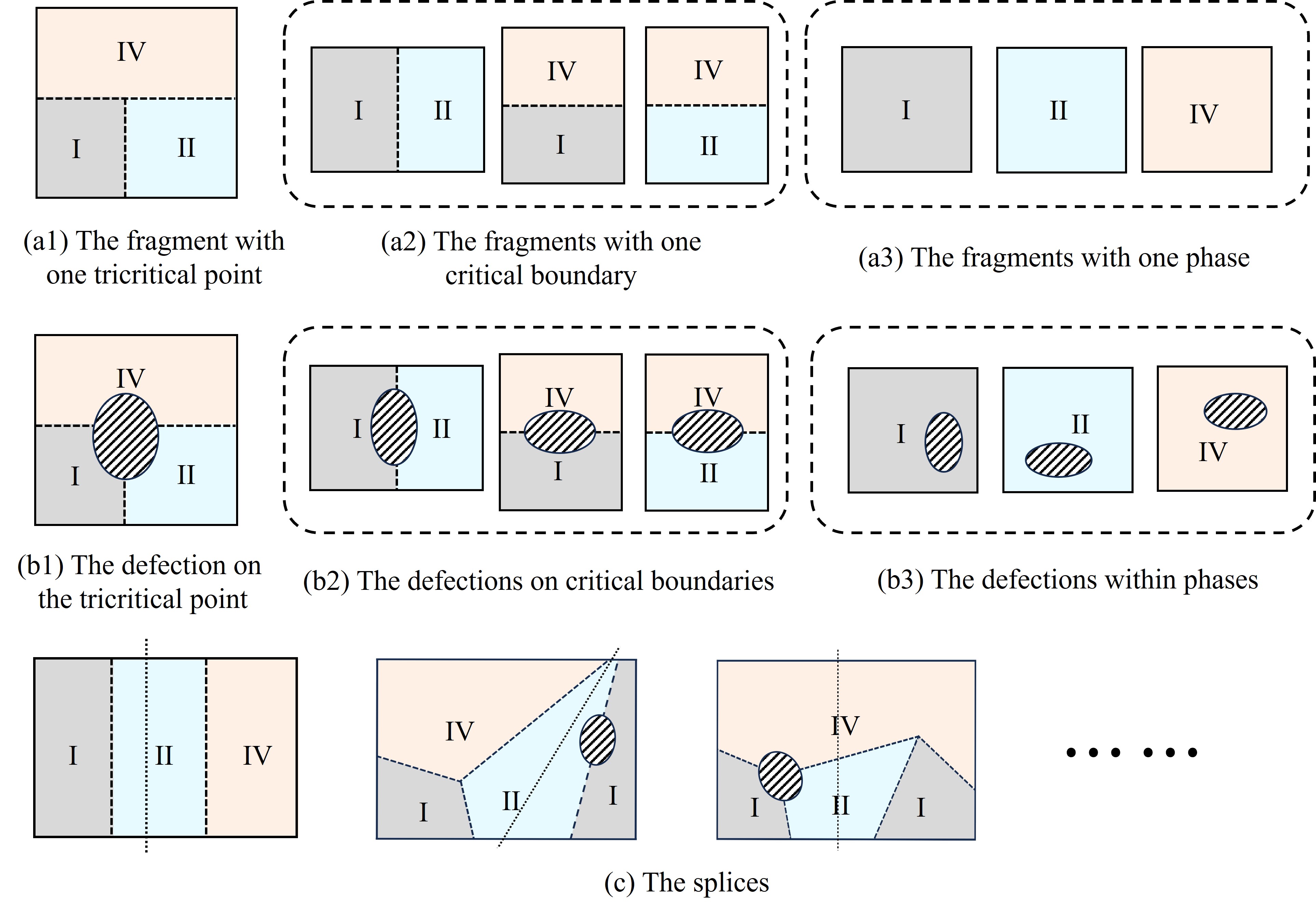}
				\caption{The three general configurations of the realistic phase diagrams. The fragments, defections and splices are listed in three rows from top to bottom.}\label{supp_real}
		\end{figure}}
		
		\section{Critical Scalings}
		
		Apart from the regularity of the phase diagram, different critical behaviors around the critical boundaries are determined by different coupling terms. We divide the possible scaling behaviors into four different kinds, giving four columns in the table of the main text. In this section, we give the detailed derivations for all these scaling behaviors. 
		
		
		\subsection{the transition from type-I to type-II}
		Firstly, we consider the typical transition case from type-I to type-II configuration where the order parameter will change from $\bar{O}_1=0$ to $\bar{O}_1\neq0$ with $\bar{O}_2=0$. Thus we can evaluate the function as
		\begin{align}
			f_1(O_1;\vec{\lambda})\equiv F(O_1,0;\vec{\lambda})=\sum_{v_1}g_{v_1,0}(\vec{\lambda})O_1^{v_1}
		\end{align}
		Consider the critical point $\vec{\lambda}_c$, satisfied that $g_{2,0}(\vec{\lambda}_c)=0$. According to the stability of type-I at the critcial point, it can be easy deduced that $v_m$ is even, which is defined as $v_m\equiv\min\left\{v_1~|~g_{v_1,0}(\vec{\lambda}_c)\neq0\right\}$.
		
		And we consider the environmental parameter just exceeds the critcial point as $\vec{\lambda_c}+\delta\vec{\lambda}$, then we have that 
		\begin{align}
			g_{v_1,0}(\vec{\lambda_c}+\delta\vec{\lambda})\sim\mathcal{O}(\left|\delta\vec{\lambda}\right|)<0,~ when~~2\le v_1<v_m
		\end{align}
		Usually, $\sim\mathcal{O}(\left|\delta\vec{\lambda}\right|)$ is valid unless some special case. We can evaluate the differentiation against $O_1$ at the new point $\vec{\lambda_c}+\delta\vec{\lambda}$, giving as 
		\begin{align}\label{expansion_exceed_critical_point}
			f'(\vec{\lambda_c}+\delta\vec{\lambda})=\sum_{v_1}v_1g_{v_1,0}(\vec{\lambda_c}+\delta\vec{\lambda})O_1^{v_1-1}=2g_{2,0}O_1+v_mg_{v_m,0}O_1^{v_m-1}+\sum_{v_1=3}^{v_m-1}v_1g_{v_1,0}O_1^{v_1-1}+\mathcal{O}(O_1^{v_m})
		\end{align}
		We can get that the new stable order point by $f'(\vec{\lambda_c}+\delta\vec{\lambda})=0$
		\begin{align}
			(\delta\bar{O}_1)^{v_m-2}=-\dfrac{2g_{2,0}}{v_mg_{v_m,0}}\sim\mathcal{O}(\left|\delta\vec{\lambda}\right|),
		\end{align}
		which is getting from the first two terms on the r.h.s. of Eq.~(\ref{expansion_exceed_critical_point}). Because the third and fourth terms of Eq.~(\ref{expansion_exceed_critical_point}) are suppressed by the first and second terms as $\left|\delta\vec{\lambda}\right|\to0$. So we can get the critical scaling law as $\delta \bar{O}_1\sim\left|\delta\vec{\lambda}\right|^{1/(v_m-2)}$. This conclusion is the same as that in single order parameter phase transition. In the critical point for 2nd order phase transition, the coefficient of the 4th order term is nonzero, i.e., $v_m=4$, which gives $1/2$ scaling rate. While at the tricritical point, where the coefficient of the 4th order term vanishes and $v_m=6$, the scaling rate will change into $1/4$ scaling rate.
		
		\subsection{the transition from type-I to type-IV in category A}
		Similar to the approach applied above, we can extend the conclusion to the two dimensional case, where the change occurs at the critical point from $\bar{O}_{1,2}=0$ to $\bar{O}_{1,2}\neq0$. 
		
		Then, we consider the expression of the free energy, giving as
		\begin{align}\label{free_energy_A}
			F=g_{2,0}(\vec{\lambda})O_1^2+g_{0,2}(\vec{\lambda})O_2^2+g_{1,1}(\vec{\lambda})O_1O_2+\bar{\sum}_{v_1,v_2}g_{v_1,v_2}(\vec{\lambda})O_1^{v_1}O_2^{v_2},
		\end{align}
		where $\bar{\sum}$ means the summation excluding the former three terms. Consider the minimum of the system is always located at $\dfrac{\partial F}{\partial O_1}=0$, which reads as
		\begin{align}\label{Eq_1st_A}
			2g_{2,0}(\vec{\lambda})O_1+g_{1,1}(\vec{\lambda})O_2+\bar{\sum}_{v_1,v_2}v_1g_{v_1,v_2}(\vec{\lambda})O_1^{v_1-1}O_2^{v_2}=0.
		\end{align}
		Around the critical boundary where $\vec{\lambda}\to\vec{\lambda_c}$, the solution $\left(\bar{O}_1,\bar{O}_2\right)\rightarrow(0,0)$, leading that the terms in $\bar{\sum}$ are higher order small contributions compared to the former two terms in Eq.~(\ref{Eq_1st_A}). It gives us a solution as
		\begin{align}\label{O_1O_2_A}
			O_1=-\dfrac{g_{1,1}(\vec{\lambda})O_2}{2g_{2,0}(\vec{\lambda})}.
		\end{align}
		Insert it back into the Eq.~\ref{free_energy_A}, we obtain
		\begin{align}
			F&=[g_{0,2}(\vec{\lambda})-\dfrac{g^2_{1,1}(\vec{\lambda})}{4g_{2,0}(\vec{\lambda})}]O_2^2+\bar{\sum}_{v_1,v_2}g_{v_1,v_2}(\vec{\lambda})[-\dfrac{g_{1,1}(\vec{\lambda})O_2}{2g_{2,0}(\vec{\lambda})}]^{v_1}O_2^{v_2}\notag\\
			&\equiv D(\vec{\lambda})O_2^2+\bar{\sum}_{v_1,v_2}g_{v_1,v_2}(\vec{\lambda})[-\dfrac{g_{1,1}(\vec{\lambda})}{2g_{2,0}(\vec{\lambda})}]^{v_1}O_2^{v_1+v_2}.
		\end{align}
		In the second line, we define that $D(\vec{\lambda})\equiv g_{0,2}(\vec{\lambda})-\dfrac{g^2_{1,1}(\vec{\lambda})}{4g_{2,0}(\vec{\lambda})}$. The condition $D(\vec{\lambda})=0$ gives nothing but the critical boundary as Eq~(\ref{critical_boundary_A}). While the minimum $v_1+v_2$ decides the scaling rate of $O_2$ around the critical boundary $\vec{\lambda}\to\vec{\lambda_c}$, writes
		\begin{align}
			\delta\bar{O}_2\sim\left|\delta\vec{\lambda}\right|^{1/(v_m-2)},
		\end{align}
		where $v_m$ is defined as $v_m\equiv\min\left\{v_1+v_2~\Bigg|~g_{v_1,v_2}(\vec{\lambda_c})\neq0,~v_1+v_2>2\right\}$. Consider the Eq.~\ref{O_1O_2_A}, we have
		\begin{align}
			\delta\bar{O}_1\sim\delta\bar{O}_2\sim\left|\delta\vec{\lambda}\right|^{1/(v_m-2)}.
		\end{align}
		
		
		
		\subsection{the transition from type-II to type-IV}
		In this case, the critical boundary is given by $C_{22}(\bar{O}_1,0;\vec{\lambda}_c)=0$, and the $\bar{O}_1$ keeps nonzero while the $\bar{O}_2$ will change from zero to nonzero.
		\begin{align}\label{delta_F0}
			\delta F\equiv F(O_1,O_2;\vec{\lambda})-F(\bar{O}_1,0;\vec{\lambda})=\sum_{i}A_i(\bar{O}_1,0;\vec{\lambda})\delta O_i+\dfrac{1}{2}\sum_{i,j}C_{ij}(\bar{O}_1,0;\vec{\lambda})\delta O_i\delta O_j+\mathcal{O}[(\delta O_i)^3].
		\end{align}
		Consider that the $(\partial_1)^n\partial_2F=0,~\forall n$, then we can calculate the minimum value $\left(\delta \bar{O}_1,\delta \bar{O}_2\right)$ by asking $\dfrac{\partial\delta F}{\delta\delta O_i}=0$ as follows
		
		\begin{align}\label{solve_min1}
			\begin{cases}
				0=\dfrac{\partial\delta F}{\partial\delta O_1}=A_1(\vec{\lambda}_c+\delta\vec{\lambda})+C_{11}(\vec{\lambda}_c+\delta\vec{\lambda})\delta O_1+\dfrac{\partial_1\partial_2^{v_{m1}}F(\vec{\lambda}_c+\delta\vec{\lambda})(\delta O_2)^{v_{m1}}}{v_{m1}!}+\mathcal{O}[\dots]\\
				0=\dfrac{\partial\delta F}{\partial\delta O_2}=C_{22}(\vec{\lambda}_c+\delta\vec{\lambda})\delta O_2+\partial_1\partial_2^{v_{m1}}F(\vec{\lambda}_c+\delta\vec{\lambda})\delta O_1(\delta O_2)^{v_{m1}-1}+\dfrac{\partial_2^{v_{m2}}F(\vec{\lambda}_c+\delta\vec{\lambda})(\delta O_2)^{v_{m2}-1}}{(v_{m2}-1)!}+\mathcal{O}[\dots]
			\end{cases}.
		\end{align}
		Here we define $v_{m1}$ and $v_{m_2}$ as $v_{m1}\equiv\min\left\{v~|~\sum_{v_1}v_1\bar{O}_1^{v_1-1}g_{v_1,v}(\vec{\lambda}_c)\neq0\right\}$, and $v_{m2}\equiv\min\left\{v~|~\sum_{v_1}\bar{O}_1^{v_1}g_{v_1,v}(\vec{\lambda}_c)\neq0\right\}$. According to the stability at $(\bar{O}_1,0)$, we can expand the $A_1$ and $C_{22}$ at $\lambda_c$ as $A_1(\vec{\lambda}_c+\delta\vec{\lambda})\sim\mathcal{O}(\left|\delta\vec{\lambda}\right|)$ and $C_{22}(\vec{\lambda}_c+\delta\vec{\lambda})\sim\mathcal{O}(\left|\delta\vec{\lambda}\right|)<0$. This equation gives us the scaling rate as 
		\begin{align}
			\delta \bar{O}_1\sim\left|\delta\vec{\lambda}\right|^{v_{m1}/(2v_{m1}-2)},~~\delta \bar{O}_2\sim\left|\delta\vec{\lambda}\right|^{1/(2v_{m1}-2)},~~if:~2v_{m1}\le v_{m2}\\
			\delta \bar{O}_1\sim\left|\delta\vec{\lambda}\right|^{\min\left\{v_{m1}/(v_{m2}-2),1\right\}},~~\delta \bar{O}_2\sim\left|\delta\vec{\lambda}\right|^{1/(v_{m2}-2)},~~if:~2v_{m1}>v_{m2}
		\end{align}
		
		We take a case as $v_{m1}=2$ and $v_{m2}=4$, thus we give
		
		\begin{align}\label{solve_min0}
			\begin{cases}
				0=\dfrac{\partial\delta F}{\partial\delta O_1}=A_1(\vec{\lambda}_c+\delta\vec{\lambda})+C_{11}(\vec{\lambda}_c+\delta\vec{\lambda})\delta O_1+\dfrac{1}{2}\partial_1\partial_2^2F(\vec{\lambda}_c+\delta\vec{\lambda})(\delta O_2)^2+\mathcal{O}[\dots]\\
				0=\dfrac{\partial\delta F}{\partial\delta O_2}=C_{22}(\vec{\lambda}_c+\delta\vec{\lambda})\delta O_2+\partial_1\partial_2^2F(\vec{\lambda}_c+\delta\vec{\lambda})\delta O_1\delta O_2+\dfrac{1}{6}\partial_2^4F(\vec{\lambda}_c+\delta\vec{\lambda})(\delta O_2)^3+\mathcal{O}[\dots]
			\end{cases}.
		\end{align}
		We can see the $\delta O_1,(\delta O_2)^2\sim\mathcal{O}(\left|\delta\vec{\lambda}\right|)$. In a special case where $A_1(\vec{\lambda}_c+\delta\vec{\lambda})=0$ with a well selected direction of $\delta\vec{\lambda}$ (Usually, it's valid if we select $\lambda_1$ and $\lambda_2$ to control the phase transtion described by the order parameter $O_1$ and $O_2$, respectively.), then we can get that 
		\begin{align}
			\delta \bar{O}_1\approx-\dfrac{\partial_1\partial_2^2F(\bar{O}_1,0;\vec{\lambda}_c)}{2C_{11}(\bar{O}_1,0;\vec{\lambda}_c)}(\delta \bar{O}_2)^2.
		\end{align}
		Because the denominator is positive according to the stability, we can see that the sign of the term $\partial_1\partial_2^2F(\bar{O}_1,0;\vec{\lambda}_c)$ can decide whether the $\bar{O}_1$ increases or decreases after the phase transition. {\color{red}This is one of the main results of the text.} With the condition $A_1(\vec{\lambda}_c+\delta\vec{\lambda})=0$ unchanged, we have
		\begin{align}
			\delta \bar{O}_1\approx-\dfrac{\partial_1\partial_2^{v_{m1}}F(\bar{O}_1,0;\vec{\lambda}_c)}{v_{m1}!C_{11}(\bar{O}_1,0;\vec{\lambda}_c)}(\delta \bar{O}_2)^{v_{m1}}.
		\end{align}
		for the general $v_{m1}$ and $v_{m2}$ case. the sign of the term $\partial_1\partial_2^{v_{m1}}F(\bar{O}_1,0;\vec{\lambda}_c)$ can decide whether the $\bar{O}_1$ increases or decreases after the phase transition.
		
		\subsection{the transition from type-I to type-IV in Category B and Category $A_2$}
		In this case, the critical condition is given by $C_{22}(0,0;\vec{\lambda}_c)=g_{0,2}(\vec{\lambda}_c)=0$. In category B, it can be viewed as the special case of the last section with $\bar{O}_1=0$ and $A_1(\vec{\lambda}_c+\delta\vec{\lambda})=0$, thus we can get the similar resluts as 
		\begin{align}
			\delta \bar{O}_1\sim\left|\delta\vec{\lambda}\right|^{v_{m1}/(2v_{m1}-2)},~~\delta \bar{O}_2\sim\left|\delta\vec{\lambda}\right|^{1/(2v_{m1}-2)},~~if:~2v_{m1}\le v_{m2}\\
			\delta \bar{O}_1\sim\left|\delta\vec{\lambda}\right|^{v_{m1}/(v_{m2}-2)},~~\delta \bar{O}_2\sim\left|\delta\vec{\lambda}\right|^{1/(v_{m2}-2)},~~if:~2v_{m1}>v_{m2}
		\end{align}
		And we have the relation
		\begin{align}
			\delta \bar{O}_1\approx-\dfrac{\partial_1\partial_2^{v_{m1}}F(0,0;\vec{\lambda}_c)}{v_{m1}!C_{11}(0,0;\vec{\lambda}_c)}(\delta \bar{O}_2)^{v_{m1}}.
		\end{align}
		same as above, due to that $A_1(\vec{\lambda}_c+\delta\vec{\lambda})=0$ is always valid here. Here we define $v_{m1}$ and $v_{m2}$ as $v_{m1}\equiv\min\left\{v~|~g_{1,v}(\vec{\lambda}_c)\neq0\right\}$, and $v_{m2}\equiv\min\left\{v~|~g_{0,v}(\vec{\lambda}_c)\neq0\right\}$.
		
		While in Category $A_2$ if the critical boundary is brunch $l_{13}$, then the critical behaviour is the same as that in categroy B. While if the critical boundary is brunch $l_{12}$, then the critical behaviour is the counterpart whose subscript $1$ and $2$ are exchanged, And the definitions of $v_{m1}$ and $v_{m_2}$ are also changed as $v_{m1}\equiv\min\left\{v~|~g_{v,1}(\vec{\lambda}_c)\neq0\right\}$, and $v_{m2}\equiv\min\left\{v~|~g_{v,0}(\vec{\lambda}_c)\neq0\right\}$. Therefore, we can see that the critical behaviour is absolutely different from the in category A.
		
	\end{document}